\documentclass[conference]{IEEEtran}
\IEEEoverridecommandlockouts
\usepackage{cite}
\usepackage{amsmath,amssymb,amsfonts}
\usepackage{algorithmic}
\usepackage{graphicx}
\usepackage{textcomp}
\usepackage{xcolor}
\usepackage{wrapfig}
\usepackage{color,soul}
\usepackage{multirow}
\usepackage{url}
\usepackage{tikz}

\usepackage[linesnumbered,ruled,vlined]{algorithm2e}
\usepackage{algorithmic}
\newcommand{\model}{\emph{\textit{CrowdSim}}}

\title{\model: A Hybrid Simulation Model for Failure Prediction in Crowdsourced Software Development}

\author{\IEEEauthorblockN{Razieh Saremi, Ye Yang \\and Gregg Vesonder}
\IEEEauthorblockA{School of Systems and Enterprices\\
Stevens Institute of Technology\\
Hoboken, NJ, USA\\
\{rsaremi, yyang4, gvesonde\}@stevens.edu}
\and
\IEEEauthorblockN{Guenther Ruhe}
\IEEEauthorblockA{Software Engineering Decision Support Lab\\
University of Calgary\\ Alberta, Canada\\
ruhe@ucalgary.ca}
\and
\IEEEauthorblockN{He Zhang}
\IEEEauthorblockA{Software Institute\\ Nanjing University\\ Jiangsu, China\\
hezhang@nju.edu.cn}}

\begin{document}
\newpage
\setcounter{page}{1}

\maketitle

\bigbreak

\begin{abstract}

A typical crowdsourcing software development (CSD) marketplace consists of a list of software tasks as service demands and a pool of freelancer developers as service suppliers. Highly dynamic and competitive CSD marketplaces may result in task failure due to unforeseen risks, such as increased competition over shared worker supply, or uncertainty associated with workers' experience and skills, and so on. To improve CSD effectiveness, it is essential to better understand and plan with respect to dynamic worker characteristics and risks associated with CSD processes. In this paper, we present a hybrid simulation model, {\model}, to forecast crowdsourcing task failure risk in competitive CSD platforms. {\model} is composed of three layered components: the macro-level reflects the overall crowdsourcing platform based on system dynamics, the meso-level represents the task life cycle based on discrete event simulation, and the micro-level models the crowd workers’ decision-making processes based on agent-based simulation. {\model} is evaluated through three CSD decision scenarios to demonstrate its effectiveness, using a real-world historical dataset, and the results demonstrate {\model}'s potential in empowering crowdsourcing managers to explore crowdsourcing outcomes with respect to different task scheduling options.

\end{abstract}
\bigbreak

\bigbreak 
\noindent\textit{This work has been submitted to the IEEE for possible publication. Copyright may be transferred without notice, after which this version may no longer be accessible.}
\bigbreak

\section{Introduction}

As an emerging paradigm, Crowdsourced Software Development (CSD) has been increasingly adopted to develop software applications \cite{lakhani2010topcoder}\cite{mao2017survey}. A crowdsourced environment happens where there is a pool of undefined human resources with internet access engaged in an articulated market of online open call projects \cite{howe2008crowdsourcing}. The CSD platform can be considered as a supply-and-demand market between tasks providers and workers \cite{difallah2016scheduling}. In order to function effectively, it must address both the needs of task providers as demands and crowd workers as suppliers. From task owners perspective, a CSD platform needs to ensure their tasks receiving broad participation, and on-time, high-quality deliveries. From crowd workers' perspective, a CSD platform needs to address a diverse set of worker motivations for participation \cite{yang2015award}\cite{karim2016decision}, such as winning task prizes, gaining community reputation, building technical skills, and so on. 
%it is very important to understand the CSD market in terms of the pool of available tasks in the platform, the similarity among available tasks in the pool and the new arrival tasks, the amount of active crowd workers, as well as crowd workers’ skill sets and performance history. While a crowd worker is more interested in required skill set for performing the task \cite{difallah2016scheduling} \cite{faradani2011s}, competition level in terms of other opponent in the competition and probability of winning the competition \cite{archak2010money}. 

In CSD practice, any mismatch between these needs may lead to task failure. Task failure happens due to lack of  available skillful workers\cite{archak2010money}, or crowd workers who are not interested in taking the tasks based on some personal
utility algorithm, their skills and some unknown factors\cite{faradani2011s}.
Existing studies reveal a 82.9\% task-dropping rate in software
crowdsourcing marketplace, which leads to a 15.7\% of crowdsourcing failure\cite{yang2016should}.
%In order to achieve a low task failure rate, a CSD market must address a worker’s need for a clearly presented and achievable task, but also ensure a task owner’s need for broad participation, on-time, and high-quality completion.
% For a task owner, it is very important to understand the CSD market in terms of the pool of available tasks in the platform, the similarity among available tasks in the pool and the new arrival tasks, the amount of active crowd workers, as well as crowd workers’ skill sets and performance history. While a crowd worker is more interested in required skill set for performing the task \cite{difallah2016scheduling} \cite{faradani2011s}, competition level in terms of other opponent in the competition and probability of winning the competition \cite{archak2010money}. 
On the one hand, task owners do not know or have control over crowd workers, which raises concerns on the qualification and trustworthiness of the crowd workers \cite{khanfor2017failure, mao2017survey, karim2016decision}. On the other hand, crowd workers are shared resources frequently multi-tasking on different requesters tasks based on their individual goals and preference. These implicit factors frequently complicate the CSD processes and cause resource discrepancy, and consequently increase the likelihood of task failure. %Hence, it is important to understand crowd workers’ sensitivity to the arrival tasks in the platform in order to attract enough reliable workers to perform tasks, return qualified submissions, and as a result reduce the task failure ratio in the platform.
Existing literature shows that workers are more interested in tasks matching their expertise and/or prior experience, and their preference on task selection %\cite{surowiecki2005wisdom} 
\cite{yang2015award} \cite{yuen2012task}\cite{yin2020matchmaker}. 
%It seems that attracting workers to a group of similar tasks may cause zero registration for lower similar available tasks, not receiving submissions or qualified submissions for registered tasks due to lack of time. 
Generally speaking, task similarity is one of the most important factors in a worker’s decision-making process, which is based on a worker’s individual knowledge and skill set \cite{brabham2013crowdsourcing}. Higher task similarity provides more choices for crowd workers to choose tasks, and higher openness for platforms to work with different task providers. As a platform moved from low to moderate levels of openness, innovation among crowd workers increased \cite{eisenmann2009opening} \cite{boudreau2008opening}.

% It is reported that the task description and competition level per task \cite{khanfor2017failure} are the most influencing factors for workers’ performances, which is associated with task failure in CSD . 
Task scheduling is a common method to prevent resource discrepancy, and decrease task failure \cite{difallah2016scheduling}. Various task scheduling methods, such as HIT-Bundle \cite{difallah2016scheduling} and QOS \cite{khazankin2011qos}, have been introduced in the crowdsourcing research field to aid the process of researching, designing, developing and training system to reduce task failures. 
Existing scheduling methods primarily focus on static aspects of tasks, such as %task status, 
task ID, or task's type, or task's required technology and neglect dynamic aspects of tasks, such as workers’ registration on last T days per similar tasks or workers’ average submissions rate on similar arrival tasks or number  in the platform. %To reflect the dynamic aspect of scheduling, there is a need to focus on impact of workers’ decision-making in CSD platform. 
In addition, existing scheduling methods fail to provide enough information to predict different task execution patterns based on workers' decision making. %Therefore, it is helpful to provide deeper insights on factors that attract reliable crowd workers and reduce the task failure ratio in CSD. 
To that end, this paper proposes a novel hybrid simulation model, {\model} for task failure prediction in CSD. It integrates three different simulation techniques, i.e. system dynamics simulation (SDS)\cite{forrester1961industrial}, discrete event simulation (DES)\cite{gordon1961general}, and agent-based simulation (ABS) models\cite{bonabeau2002agent} , to provide insights on the effects of various compounding factors \cite{zhang2011impact} \cite{zhang2012toward} for attracting reliable crowd workers on the task failure risk in CSD. While SDS methods represent the overall view of the system and dynamic relation among the stakeholders \cite{forrester1961industrial}, DES models capture the actual process-level details in development \cite{gordon1961general}, and ABS models simulate the behavior among individual workers and social groups \cite{bonabeau2002agent}.  Moreover, simulation allows one to reenact a scenario in order to identify bottlenecks in the process and provide solutions \cite{angelopoulou2015simulation}. 

%In developing the hybrid simulation model, we leverage empirical findings including crowd worker behavior, task completion outcome, and crowd team performance from our previous works \cite{yang2015award}\cite{saremi2015empirical}\cite{saremi2017leveraging}, and packaged knowledge on SDS, DES, and ABS simulation models in different layers as three different components of the proposed simulation model. SDS model simulation follows empirical finding on workers behavior based on task speciation \cite{yang2015award}, DES model simulates task completion flow based on task arrival patterns in CSD \cite{saremi2015empirical} and ABS model simulates workers performance following empirical findings of team performance \cite{saremi2017leveraging}. 

{\model}aims at answering the following question:
\textbf{Given a competition context in a CSD marketplace, how to strategically schedule a new task to ensure task success and prevent task failure?} 
More specifically, this paper reports the design and impact of three different components of {\model} in the context of crowdsourcing task scheduling.  In designing {\model}, we leverage empirical findings including crowd worker behavior, task completion outcome, and crowd team performance from our previous works \cite{yang2015award}\cite{saremi2015empirical}\cite{saremi2017leveraging}, and packaged knowledge on SDS, DES, and ABS simulation models in different layers as three different components of the proposed simulation model. SDS model simulation follows empirical finding on workers behavior based on task speciation \cite{yang2015award}, DES model simulates task completion flow based on task arrival patterns in CSD \cite{saremi2015empirical} and ABS model simulates workers performance following empirical findings of team performance \cite{saremi2017leveraging}. {\model} is evaluated through three CSD decision scenarios to demonstrate its effectiveness, using a real-world historical dataset, and the results demonstrate {\model}'s potential in empowering crowdsourcing managers to explore crowdsourcing outcomes with respect to different task scheduling options. To the best of our knowledge, there is no existing study using a hybrid simulation model of CSD based on a combination of all three different simulation methods of SDS, DES, and ABS.

The rest of the paper is organized as follows: Section 2 presents a motivational example in CSD task scheduling, Section 3 provides the overview of the proposed work, Section 4 presents the implementation of the hybrid simulation model, Section 5 illustrates its effectiveness through two case studies, Section 6 is related work, and Section 7 concludes the paper with future work.  

\section{Background}

\subsection{A Motivational Example}

\begin{figure*}
\centering

\includegraphics[width=1\textwidth,height=0.45\textwidth, keepaspectratio]{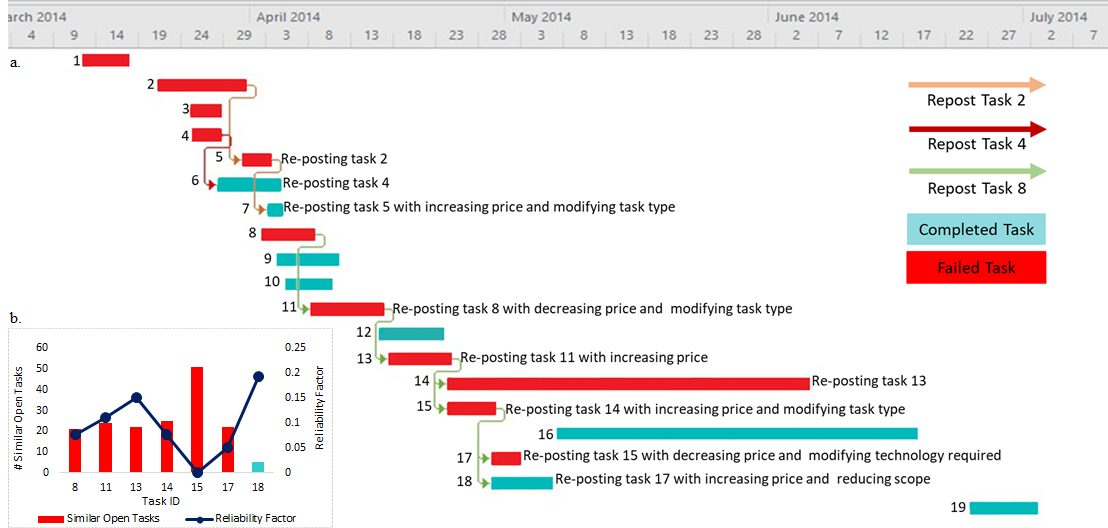}
\caption{An Example CSD Project: a. Gantt-Chart, b. Comparison of the number of simultaneously open, similar  tasks and average worker reliability across 7 re-postings}

\label{Motivation}
\end{figure*}

Figure \ref{Motivation} illustrates a CSD project on the TopCoder platform, which consists of 19 tasks with a total duration of 110 days. The project experienced a 57\% task failure ratio, meaning 11 of the 19 tasks were not successful. More specifically, %Task 14 and 15 failed due to client request, and Task 3 failed due to unclear requirements. The remaining 
8 tasks (i.e. \#1, 2, 4, 5, 8, 11, 13, and 17) failed due to receiving zero submissions, and other reasons for failure include client request and unclear requirements, etc. These tasks are basically repeated postings of three distinct tasks (i.e. tasks 2, 4, and 8) %that were re-posted after each failure.% to be successfully completed as the new task. 
As illustrated in Figure \ref{Motivation}-a, task 2 was cancelled and re-posted as tasks 5 and 7, which was finally completed on the third attempt. %When it was re-posted as Task 7 with changes in the monetary prize and task type. 
Similarly, task 4 was re-posted as task 6, and completed with no modification. More interestingly, task 8 was re-posted 6 times, as tasks 11, 13, 14, 15, 17, and 18 respectively. Each re-posting modified Task 8 in terms of the monetary prize, task type, and required technology. Finally, the task was successfully completed as Task 18.

% \begin{figure}
% \centering
% \includegraphics[width=0.9\columnwidth,keepaspectratio]{Task8.png}
% \caption{Details of Task Similarity and Reliability Factor on Journey of Task 8 }
% \label{Task8}
% \end{figure}

Would it be possible that task similarity and worker reliability play role in task success? As shown in Figure \ref{Motivation}-b, this seems to be true from the comparison of the number of simultaneously open, similar tasks as well as average worker reliability across 7 re-postings for task 8. More specifically,
these 7 re-posted tasks were competing on average with 27 other tasks with similarity level as degree of task openness greater than 60\%, while the successful re-posted tasks were only competing with five tasks with the same level of openness. Interestingly, the task with lowest number of competing open tasks attracted workers with higher average reliability factor of 0.19.
Task 8’s journey in the motivation example supports the fact that the level of task openness directly impacts task success. Also, a lower level of task openness in the platform positively impacts attracting workers with a higher reliability factor; this relationship is demonstrated in Figure \ref{Motivation}-b. Attracting workers with a higher reliability factor increases the chance of receiving qualified submissions and task success. 
Existing research reflects that crowd workers are more interested in working on tasks with similar skillset requirements \cite{difallah2016scheduling}. Compared with general crowdsourcing platforms, CSD platforms are designed for crowdsourcing competitively paid, complex, and knowledge-intensive software development tasks.  Therefore, it is assumed that there is a much stronger correlation between the degree of a task's openness in terms of task similarity and resource diversity than in terms of skillset. However, observations from the motivational example indicate that it is difficult to draw inclusive conclusions regarding the causality relationships among task attributes, worker attributes, and task failure, due to the complicated, intertwining relationships among them. %To the best of our knowledge, none of the available studies focused on the causal similarity among simultaneously open tasks on the platform and its impact on workers’ availability and task failure. 
Therefore, providing deeper insight into the relationship between worker reliability and task openness is a missing factor that can help project managers’ decision-making process in scheduling tasks.

\subsection{Previous Work}

%In general, new task postings and worker arrivals are discrete events, typically following exponential distribution \cite{saremi2015dynamic}. In addition, the higher the level of diversity among active workers, the lower the degree of task failure \cite{saremi2015dynamic}. 
Our previous empirical studies led to the understanding and development of empirically derived worker behavior patterns, open team performance patterns, and task completion patterns. The results suggest that:
\textit{\textbf{(1) Crowd worker behaviors},} in general, have an inverse U-shape relation with the monetary prize associated with task\cite{yang2015award}. In addition, there is a direct relationship between the competition level per task and number of submissions; however, increasing the level of task attractiveness for workers leads to a decrease in the number of submissions in some cases. Interestingly, workers’ performance follows a variety of U-shape curves by changing task openness for similar tasks \cite{yang2015award}. \textit{\textbf{(2) Crowd team performance}} indicates high responsiveness but low submission potential. More than half of workers responded to a task call on the very first day of task’s publication and registered for them, while only 24\% of registrants make a submission (i.e. 14\% of available workers) \cite{saremi2017leveraging}. Furthermore, about 76\% of the submitters exceeded the acceptance criteria, i.e. only accounting for  10\% of available workers \cite{saremi2017leveraging}. More specifically, higher experience among workers resulted in a higher chance of task submissions (i.e. 60\%), while lower experience resulted in a lower level of task submissions (i.e. 25\%); average experience resulted in a submissions level of around 57\% \cite{saremi2017leveraging}. Table \ref{W.Performance} summarizes workers’ performance on the platform.
\textit{\textbf{(3) Task completion}} signifies high potential for schedule acceleration. Due to mini-task parallelism %is the strongest method of task arrival in CSD
\cite{saremi2015empirical}, CSD is associated with an 82\% reduction in the schedule acceleration rate \cite{saremi2017leveraging}. A higher degree of parallelism can lead to  increased task competition and a shorter planning schedule to complete the project, thus providing consequently better resource allocation. %Moreover, it is reported that about 75\% of posted tasks would price under 67\% of total project budget in parallel mass production \cite{saremi2015empirical}.

%\begin{small}
\begin{table}[!ht]
\caption{Workers’ Performance in the CSD Platform} % title of Table
\centering % used for centering table
\begin{tabular}{p{2.25cm} p{0.7cm} p{2cm} p{1cm} p{0.7cm}}
\hline

Workers' Belt & & Rating Range(X) &  Workers\% & p(qualified Sub)\\ %[0.5ex] % inserts table
\hline%\hline % inserts single horizontal line

\multirow{1}{*}{Lower Experienced} 
& Gray &  $ {X < 900} $ & 90.02\% & 0.25 \\

\hline

\multirow{3}{*}{Average  Experienced} 
& Green & ${900 < X < 1200}$ & 2.88\% & 0.45 \\

& Blue & ${1200 < X < 1500}$ & 5.39\% & 0.39 \\

\hline

\multirow{3}{*}{Higher Experienced} 
& Yellow & ${1500 < X < 2200}$ & 1.54\% & 0.6\\

& Red & ${X > 2200 }$ & 0.16\% & 0.6 \\

\hline
\label{W.Performance}
\end{tabular}
\end{table}

%\end{small}
\section{The Proposed \model  Model}\label{Proposed}
\subsection{Definitions }

In this section, we first define the basic notions and definitions used in {\model}, involving four types of entities: platform, tasks, projects, and crowd-workers. %Details of each entity are described at below.

\textbf{\textit{Platform}}:  A platform is a marketplace where companies place open calls for specific software development tasks to be completed by a large pool of undefined groups of external software workers\cite{stol2014two}. The general purpose of a crowdsourcing platform is to provide a market in which a company can: (1) advertise tasks to be crowdsourced; (2) specify task rewards paid to task winners; and (3) allow crowd workers as agents to participate in and complete tasks\cite{fuxman2008using}. 
Therefore, at any given time t (${t_{q}}$), the platform (${Plat}$) contains a set of m projects (${Proj}$), a set of n active tasks (${Task}$), and a set of r crowd workers (${Agent}$). Note that one project might corresponds to a set of tasks, which is a subset of (${T}$). The Platform, at time $t$, is defined as follows:
\[
{Plat} = {\{t, \\Proj_t\\, \\Agent_t\\, \\Task_t\}\}}
\]
%\[
%where
%	\begin{cases}
%        \text{k= 1, 2, 3, …, m} \\
%        \text{z= 1, 2, 3, …, r} \\
%        \text{i= 1, 2, 3, …, n} \\
%        \text{q= 1, 2, 3, …, l} 
%    \end{cases}
%\]

\textbf{\textit{Task}}: CSD tasks are typically time-dependent and complex, with high inter-dependencies. In {\model}, a task (${T_{i}}$) contains a tuple of different characteristics including the task’s identification (${id_{i}}$), arrival time (${arr_{i}}$), duration (${d_{i}}$), associated award (${aw_{i}}$), requirements (${req_{i}}$), type  (${type_{i}}$), required technology  (${tech_{i}}$), as well as the 
%task’s dependency  (${TD_{i}}$) to other tasks (as series of parallel  (${P_{i}}$) or sequential tasks  (${S_{i}}$)), and 
the task’s status  (${status_{i}}$) (i.e., "completed" or "cancelled"). The CSD task set at time $t$, i.e., ${T_{t}}$  is represented as follows: 
\[
{Task_{t}} = {\{T_i\}} = {\{({id_{i}}, {d_{i}}, {arr_{i}}, {aw_{i}}, {req_{i}}, {type_{i}}, {tech_{i}}, {status_{i}})\}} 
\]
\[
where
	\begin{cases}
        %{TD_{i}} = \begin{cases}
        %            1 &{P_{i}} = 1 \\
        %            0 & {S_{i}} = 1
        %           \end{cases} \\
        {status_{i}} = \begin{cases}
                    1 &{status} = "completed" \\
                    0 & {status} = "cancelled"
                    \end{cases} \\
        \text{i = 1, 2, 3, …, n}
    \end{cases}
\]

\textbf{\textit{Project}}: A CSD project (${P_k}$) typically contains a set of time dependent tasks to be completed. %In this study, the time sequence of an arriving task is considered as a measurement of the series of parallel or sequential tasks that followed finish-to-start task dependency techniques. %Task (n), ${T_{n}}$, is parallel with task (n-1), ${T_{n-1}}$, if it starts before ${T_{n-1}}$. And ${T_{n}}$ is a sequential task for ${T_{n-1}}$ if it starts after ${T_{n}}$ is completed. 
The project duration (${pd_{k}}$) in {\model}is defined as the difference of the last sequential task's submission date (${TsSq_{n}}$) and the first sequential task's registration date (${TrSq_{1}}$). Also, the ratio of failed-to-successful tasks per project is a measure of project success. Therefore, in this research, a project is defined as a tuple of different characteristics of decomposed tasks $\{{T_{i}}\}$, project duration (${pd_{K}}$), project failure ratio (${pf_{K}}$), and the number of crowd workers (${agent_{r}}$) participated in the tasks using the following formula: 
\[
{P_{k}} = {(\{{T_{i}}\}, {pd_{k}}, {pf_{k}}, \{{agent_{r}}\})}
\]
\[
where
	\begin{cases}
        % {pd_{k}} = {\sum_{i=0}^{k}{Sq_{i}}}  \\
        {pd_{k}} = {TsSq_{n}} - {TrSq_{1}} \\ 
        {pf_{k}} = \frac{\sum_{i=0}^{k}{F_{i}}} 
                    {\sum_{i=0}^{K}{T_{i}}}  \\
       \text{k = 1, 2, 3, …, m} 
    \end{cases}
\]

\textbf{\textit{Crowd Workers}}: %According to Howe \cite{howe2008crowdsourcing}, crowd workers (${AZ}$) are a large and undefined group of skilled workers who have access to a task via the internet. In this research, 
CSD workers are a tuple of different characteristics of workers’ identification (${AID_{r}}$), reliability factor (${Re_{r}}$), rating (${Ra_{r}}$), skillset (${SK_{r}}$), score (${So_{i}}$), number of wins (${Wi_{i}}$), location (${L_{r}}$), and his/her membership tenure (${MA_{r}}$) at a given time t (${t_{q}}$). A crowd worker is defined with the following formula: 
\[
{Agent_{z}} = {({AID_{r}}, {Re_{r}}, {Ra_{r}}, {SK_{r}}, {So_{i}}, {Wi_{i}}, {L_{r}}, {MA_{r}})}
\]
\[
where
   \text{  z = 1, 2, 3, …, r} 
\]

\subsection{Overview of {\model}}\label{Hybrid}
%YY: omitted due to overly repeating.
%A crowdsourcing platform is a dynamic market that contains three direct stakeholders: a software worker as a supply agent, a demand company as a requester, and the crowdsourced platform as the market \cite{brabham2013crowdsourcing}.

\begin{figure*}[ht!]
\centering
\includegraphics[width=1\textwidth,height=0.4\textwidth]{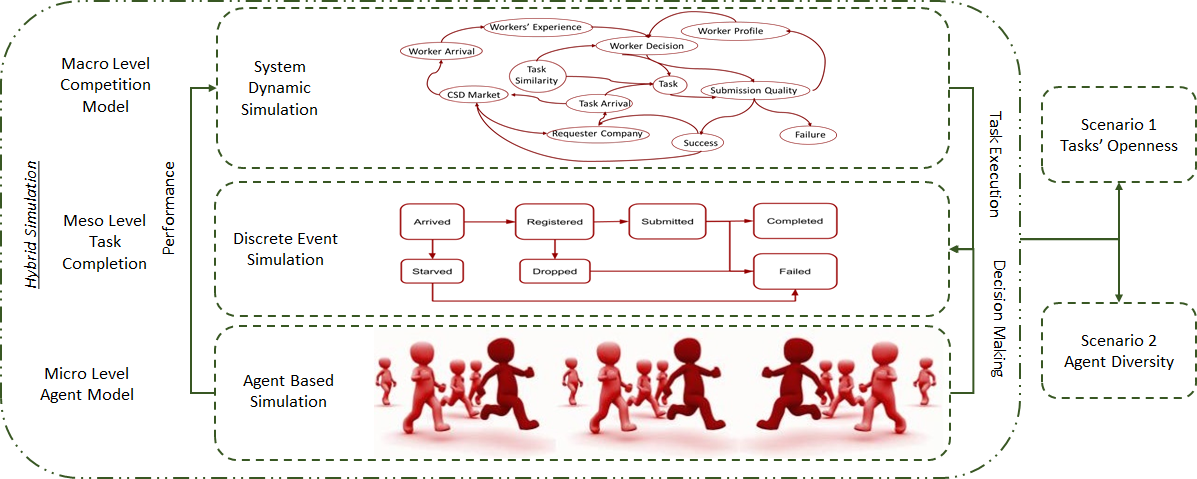}
\caption{Overview of {\model}}
\label{overview}
\end{figure*}

%Crowdsourcing’s success is based on all the individual stakeholder’s behaviors and their interactions with each other. Requestors provide the platform with tasks and software workers perform the tasks. Such platforms need to be designed for software workers’ to easily understand crowdsourcing tasks and allow them to form relationships and practical communication channels with requestors \cite{dellarocas2001analyzing}. The agent-based simulation (ABS) model is responsible for an individual agent’s behavior, while the discrete event simulation (DES) manages tasks sequentially and the system dynamic simulation (SDS) displays the interactions among the system’s parameters and feedback within the platform. 

%In the hybrid model, the DES is responsible for executing tasks in the platform. Once a task arrives in the platform, the agents’ decision-making process starts via the ABS model. 
Figure \ref{overview} illustrates the overview of {\model}, the hybrid simulation model. The agent-based simulation (ABS) model is responsible for an individual agent’s behavior, while the discrete event simulation (DES) manages tasks sequentially, and the system dynamic simulation (SDS) displays the interactions among the system’s parameters and feedback within the platform. Once a task arrives in the platform, the agents’ decision-making process starts via the ABS model. % The proposed model is created based on the TopCoder \footnote{ \url{https://www.topcoder.com/}} workflow and measures the task failure ratio in the platform. 
An agent’s different decision-making strategy defines his or her performance in the platform. Agents’ performance directly impacts task failure rate and the market’s trust in the platform. 

\subsubsection{\textbf{Macro-level Competition Model}}
%The SDS model is applied to represent how agents’ errors affect their reliability to return qualified tasks by revealing a pattern of broader agent behavior \cite{angelopoulou2018utasimo} \cite{gregoriades2008human}.
 
% \begin{figure}[ht!]
% \centering
% \includegraphics[width=1.0\columnwidth]{SDplatform.png}
% \caption{Overview of Platform Model (SD)}
% \label{SD}
% \end{figure}

At the top platform level, the SDS model, is applied to represent the overall relationship amng task arrival, worker reliability, and task competition outcome by revealing a pattern of broader market behavior \cite{angelopoulou2018utasimo} \cite{gregoriades2008human}. As illustrated in macro level of Figure \ref{overview}, it contains 14 variables including task, agent decision, workers’ performance, task similarity, worker profile, worker skillset, and different available crowdsourced markets. The SDS model represents the causal loops among different levels of the platform.

\textit{“Task arrival”} in the platform is an event modeled by a Poisson distribution. Any new arrival task will trigger its execution process modelled through a DES, and joins the platform’s pool of open tasks. More details on the DES model is provided in the following subsection. Each new task is associated with a similarity factor upon arrival. According to empirical analysis, a higher degree of “task similarity” among the pool of open tasks in the platform leads to higher competition level. An average task similarity of 90\% led to an average competition level of 55\%. However, the degree of task similarity does not impact the level of submissions; failures increased when the task similarity rose in the pool of open tasks. For example, an average task similarity of greater than 80\% provided a failure ratio of 20\%, while an average task similarity of 60\% provided an average failure ratio of 13\%.

As Figure \ref{overview} illustrates, having other available CSD platforms in the market impacts \textit{“workers’ arrival”} and a company’s decision to work with a specific platform. Also, \textit{“agents’ skillset”} and task similarity are the main factors impacting agents’ decision-making processes to post a specific task on the platform. An agent’s decision to \textit{“register”} and \textit{“submit”} a task provides information for \textit{“worker’s performance”}, and consequently updates the \textit{“worker’s profile”}.

Workers’ performances on the platform lead to different task statuses of success or failure. If the task is successfully completed, it counts as \textit{“complete”} and is reported to the \textit{“requestor company”}. In this model, task completion ratio (${TCR_{k}}$) is calculated based on the ratio of passing review tasks to the registered tasks.
\[
{TCR_{k}} = \frac{\sum_{i=0}^{n}{C_{i}}} 
{\sum_{i=0}^{n}{R_{i}}} 
\]
However, if the task is not qualified, it will count as a \textit{“failure”}. Task failure ratio (${TFR_{k}}$) in this model is the ratio of tasks that are registered but are dropped or do not pass peer review. Failure ration in this model calculates as follow.
\[
{TFR_{k}} = 1 - \frac{\sum_{i=0}^{n}{C_{i}}} 
{\sum_{i=0}^{n}{R_{i}}} 
\]
Table \ref{VSD} summarizes the structured attributes used in the SD model.

\subsubsection{\textbf{Task Completion}}\label{Task Completion2}
The task’s lifecycle is one of the most important factors in an agent’s decision-making process in crowdsourcing, since it represents a task’s priority and complexity. In this model, tasks are defined as a set of discrete events (DE) that have a start and end date.

An arrival task (${T_{j}}$) in this model is defined as a list of ${'w'}$ number of tasks from the same project that arrive simultaneously in the platform.
\[
{T_{j}} = {\{{T_{i}}\}}
\]
\[
where
	\begin{cases}
    	{{D_{j}}< {D_{j}} < {D_{j + 1}}}\\
        \text{j = 1, 2, 3, …, w} 
    \end{cases}
\]

Task arrival in this model, is an event modeled by the Poisson distribution of a population of tasks with a lambda equal to 87. The Messo level in Figure \ref{overview} illustrates the overview of the DSE model. When a task arrives, it becomes an \textit{“arrived”} task in the platform, meaning that it can be \textit{“registered”} by available agents to start the process. Empirical analysis revealed that arrival tasks will attract software workers at a rate of 70\%. If the task does not attract any software workers and faces zero registration, it will be marked as \textit{“starved”}. 

% \begin{figure}[ht!]
% \centering
% \includegraphics[width=1.0\columnwidth]{Tasklifestyle.png}
% \caption{Overview of the Task Lifecycle Model (DE)}
% \label{TLC}
% \end{figure}

Based on different decision-making scenarios by competing agents, the proposed model provides two different failure prediction ratios in different phases of the task lifecycle. TopCoder currently applies a heuristic-based color-coded (red, yellow, and green) predictor for a task outcome in the registration phase. The model involves three simple rules with respect to the sum of the reliability ratings of all software workers registered for a task \cite{ye2015crowd}. If the sum of reliability ratings is greater than or equal to two, the predictor produces a green label to symbolize the task’s success to the requester. If it is less than one, the predictor produces a red color which represents a likely failed task. Yellow is the result of a reliability rating in-between one and two, which represents an uncertain task status. In this model, the heuristic model used by TopCoder is extended to predict the first failure ratio in the registration phase, ${FPR_{k}}$, as illustrated below.
\[
{FPR_{k}} =
	\begin{cases}
        \frac{\sum_{i=0}^{n}{Re_{j}}*{P_{j}}}{3} & {\sum_{i=0}^{n}{Re_{j}} > 2} \\
        \frac{\sum_{i=0}^{n}{Re_{j}}*{P_{j}}}{2} & {1 < \sum_{i=0}^{n}{Re_{j}} < 2} \\
        \frac{\sum_{i=0}^{n}{Re_{j}}*{P_{j}}}{1} & {\sum_{i=0}^{n}{Re_{j}} < 1} \\
    \end{cases}
\]

In this equation, ${P_{j}}$ is the probability of zero qualified submissions by the registered agents as defined in Table \ref{W.Performance}. By the end of the task’s duration, new registered tasks may be \textit{“submitted”} by agents. If registered tasks face zero submissions, they were \textit{“dropped”}.
In this model, as soon as the first submission happens, the task submissions ratio (${TSR_{k}}$) is dynamically calculated based on the ratio of submitted to registered tasks in the platform (${TSR_{k}}$), as displayed below.
\[
{TSR_{k}} = 1 - \frac{\sum_{i=0}^{n}{S_{i}}} 
{\sum_{i=0}^{n}{R_{i}}} 
\]
Statistical correlation analysis on different drivers in a crowdsourced task indicated that task failure and task submissions ratio in the platform are directly related. Therefore, the second task failure prediction ratio (${FPS_{k}}$) can be analyzed based on a linear regression model of dynamic task submission ratio.
\[
{FPS_{k}} = {0.0473({TSR_{k}}) + 0.014} 
\]
The average CSD task duration is 16 days \cite{yang2015award}. According to this study’s empirical analysis, task duration followed a triangular distribution with a maximum of 30 days, a minimum of one day, and a mode of 16 days.

Submitted tasks must pass the peer review phase to ensure the quality of submissions. In this phase, if the quality of task is greater than or equal to 75, it is complete; if not, it has failed. If the submitted task is qualified, it is recorded as \textit{“completed'}; otherwise, the task is recorded as a \textit{“failure”}. Moreover, all the starved and dropped tasks are reported as failures. 

% Table \ref{VDE} summarizes the structured attributes used in the DES model.

% \begin{table}[!ht]
% \caption{Variable used in Discrete Event Simulation Model} % title of Table
% \centering % used for centering table
% \begin{tabular}{p{2cm}p{4cm}p{1.5cm}}

% \hline
% Metric & Definition & Model Variables\\
% \hline
% \\
% Registered (${R}$) & Number of registrants that are willing to compete on total number of tasks during a specific period of time. Range: (0, r) & tregister(int) \\
% \\
% Task Duration (${D}$) & Total available time from task registration start date to submission deadline. Range:(0, $\infty$) & duration(time)\\
% \\
% Submitted (${S}$) & Number of submissions that a task receives by its submission deadline in specific period of time. Range: (0, \# registrants) & tsubmit(int) \\
% \\
% Peer Reviewed (${PR}$) & Process of reviewing a submitted task to check the quality of submissions. Range: (0, \# registrants) & tpeer(int) \\
% \\
% Completed (${C}$) & Qualified task that has successfully passed the peer review. & tcomp(int)\\
% \\
% Re-worked ((${RW}$) & Qualified task that needs some adjustment in order to pass the peer review. & trework(int) \\
% \\
% Failed (${F}$) & Non-qualified task that has not passed peer review. & tfail(int) \\
% \\
% State & Each task has a state that declares the task situation during the submission process. & state (Arrived, Registered, Submitted, Reviewed, Complete) \\
% \hline
% \label{VDE}
% \end{tabular}
% \end{table}

\subsubsection{\textbf{Agent Model}}
\begin{figure}[ht!]
\centering
\includegraphics[width=0.8\columnwidth]{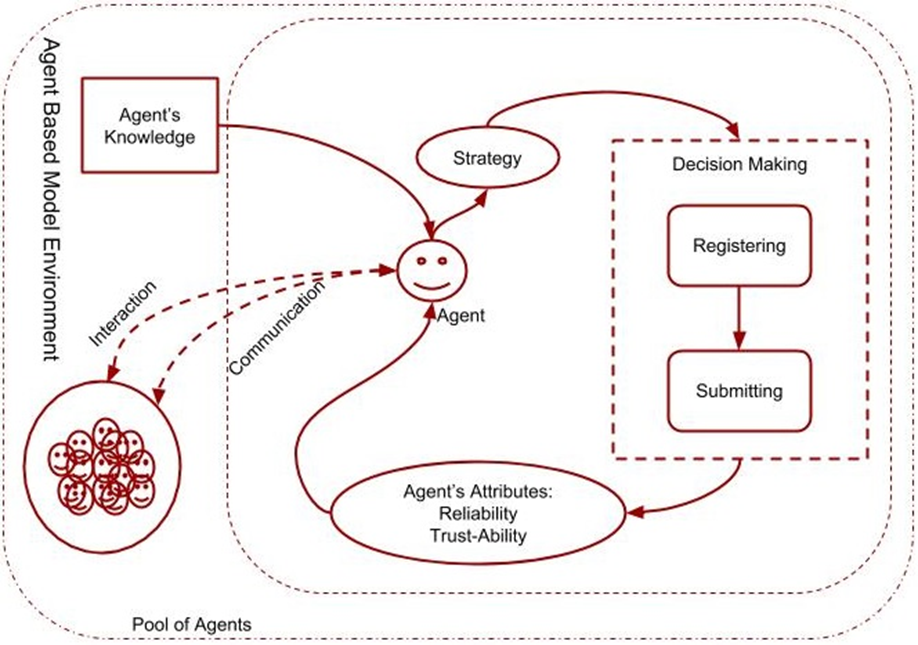}
\caption{Overview of Agent Model (AB)}
\label{AB}
\end{figure}

Crowdsourced projects integrate online and unknown workers’ elements into the design. It is reported that crowd workers often overestimate their productivity \cite{jorgensen2005over}, and they register for more tasks than they can complete. Therefore, simulating crowd-workers with various characteristics, decision-making processes, and performance ratios is difficult. Applying an agent-based (ABS) method to simulate crowd-workers’ behavior individually provides the option of observing a diversity of attributes. Crowd-workers are represented as agents who have one or more of the following characteristics:

\begin{enumerate}
    \item Identifiable with a set of rules that direct their behavior; an autonomous agent that can act independently in the environment and have control over their actions.
    \item Situated workers that work in the same environment and interacting with each other.
    \item Flexible agent that can adapt their behavior to be a better fit to the environment \cite{macal2008agent}.
\end{enumerate}
The agents’ arrival to the platform follows a non-homogenous Poisson distribution \cite{faradani2011s}. Agents are assigned unique IDs upon their entrance into the simulation. As is displayed in Figure \ref{AB}, the agent-based model contains: i) the agent environment, ii) a set of agents’ attributes, and iii) and the agent decision-making process \cite{forrester1961industrial}. In any crowdsourcing platform, an individual agent has \textit{“agent’s knowledge”}, which is based on his or her skillset, background, and the society s/he represents \cite{si2014encoding} \cite{kaufmann2011more}. The agent is also a member of the \textit{“pool of agents"}. This membership allows the agent to interact and communicate with other agents within the pool. TopCoder divides the \textit{“pool of agents”} into five groups. 

It then subdivides each of these five agent groups into five belts that correspond to skill level in the following descending order: red (highest), yellow, blue, green, and gray (lowest) \cite{yang2016should}, as it is summarized in Table \ref{W.Performance}. 

Agents update their \textit{“personal knowledge”} based on the other agents’ performances in the platform. The combination of interaction and communication with other agents in the pool and an agent’s personal knowledge creates an agent’s \textit{“strategy”} in the system.

To address the rest of the agents’ characteristics, agents’ strategy provides a utility factor \cite{difallah2016scheduling} that defines their individual behaviors. Therefore, each agent has an internal \textit{“decision-making”} process consisting of two components: \textit{“registering”} for a task and \textit{“submitting”} the task. An agent’s decision-making process is related to the information that the agent receives from the agent’s community and social environment, as well as other agent’s decision to compete on the task \cite{saremi2017leveraging}. 

Once a task is generated in the system, agents decide to register for the task. An agent’s registration for a task categorizes him or her as an active agent. Agent registration in this model follows an event at a rate of one registration per day per agent and a random number greater than 0.8. Agents perform and eventually submit a task based on a variety of factors, including the number of open tasks in their task list, a task’s complexity and competition level, as well as the probability of winning over the competition. It seems that agents analyze their probability of winning based on the task’s competition level and the number of more highly ranked opponents \cite{faradani2011s}.

It is reported that some agents use their history of victories as a policy of assuring to win the registered tasks \cite{faradani2011s}. As the result, an agent registers for a greater number of task than s/he can perform and eventually drops some of the tasks. An agent’s final decision to submit a task and his or her level of submissions impacts the agent’s profile. The agent may or may not decide to submit the registered task. A task’s submission decision is calculated at a rate of 0.51 per day per agent when the random number associated with the agent is greater than 0.86. When an agent submits a new task, an agent’s reliability factor will be updated.

According to previous research, 59\% of agents in the active pool respond to the task call in the first day, while only 24\% of the agents who initially responded to tasks call go on make a submission \cite{saremi2017leveraging}. A total of 76\% of the agents exceeded the submissions deadline \cite{saremi2017leveraging}. 48\% of registrants are among average-rated agents (i.e. those with a green or blue belt), while 86\% of higher-rated agents (i.e. those with a yellow or red belt) submit qualified submissions. The personal knowledge in the proposed ABS model reflects the reported empirical analysis. 
Also, according to empirical data, reviewed tasks with scores greater than 75\% are considered qualified and the submitting agent is reported as the winner or runner-up. Therefore, qualified submissions are determined by assigning a random number greater than 0.75 as quality score. When a submission passes peer review, an agent’s attribute rate is updated, and the new score is reported in the agent’s profile.
Decisions from simulated agents determine task progress.  The details of the model integration are explained in part appendix-\ref{finalalgorithm} algorithm 1.  Table \ref{VAB} summarizes the variables used in the agent-based model.

% \begin{table}[!ht]
% \caption{Variable Used in Agent-Based Simulation Model} % title of Table
% \centering % used for centering table
% \begin{tabular}{p{1.5cm}p{3cm}p{2cm}}

% \hline
% Metric & Definition & Model Variables\\
% \hline
% \\
% Registering (${RA}$) & Number of tasks a worker registered for in specific period of time. Range: (0, i) & a\_registering (ArrayList${<Task>}$) \\
% \\
% Submitting (${SA}$) & Number of tasks a worker submits in specific period of time. Range: (0, \#registrants] & a\_submitting (ArrayList${<Task>}$)\\
% \\
% Quality (${Q}$) & The quality of submitted task based on associated score. Range: (0, 1) & tQ(int) \\
% \\
% Reliability (${Re}$) & The percentage of successful task submissions in a worker’s most recent 15 task registrations. Range: (0, 1) & tRe(int) \\
% \\
% Trust-ability (${TA}$) & Percentage of number of qualified submissions among total number of submissions that a worker makes & tT-A(int)\\
% \hline
% \label{VAB}
% \end{tabular}
% \end{table}

\section{implementation of crowdsim}

In many cases, a simulation is a tool for decision-making that can help reduce risk on a tactical or operation level \cite{gordon1961general}, while simulating task scheduling helps on the operational level. In {\model} we extend the proposed model in \cite{saremi2015dynamic}, and create a hybrid simulation model based on SDS, DES, and ABS using Anylogic\footnote{ \url{https://www.anylogic.com/}}.  Driven by the resource-related challenges in software development, there are three steps in {\model}: the micro-, meso-, and macro-level.

%\bigbreak
\subsection{ Macro-level: Platform Competition}
Tasks and agents created by daily events enter the existing pool of open tasks and available agents in the platform, also known as the dynamic crowdsourced marketplace. Task execution follows the project schedule provided by DES. Each task is associated with a similarity rate created by an event that follows a uniform distribution between 30\% and 98\% based on our previous empirical analysis \cite{10.1007/978-3-030-50017-7_7}. In this model, agent arrivals follow the Poisson distribution. Arriving tasks in the platform impact an agent’s decision-making process and consequently, the agent’s experience. An agent’s experience in the model follows beta distribution with a minimum of zero and a maximum of 3000 where $\alpha = 1$ and $\beta$ = 5. The result of the SDS model provides the platform failure ratio. 

%\bigbreak
\subsection{Meso-level: Task Completion}
   
\begin{figure}[ht!]
\centering
\includegraphics[width=130pt,height=13pc]{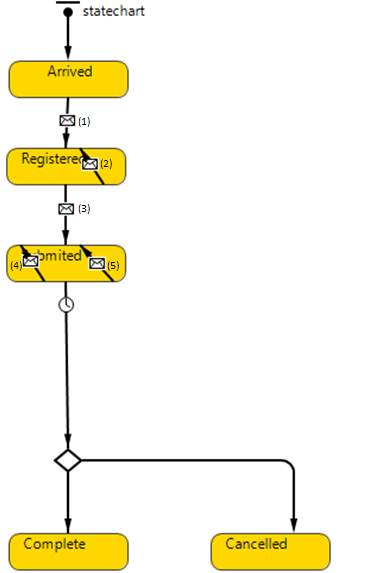}
\caption{State Chart, Task Completion Model}
\label{ADE}
\end{figure}

Task arrival in the DES model is sequential and follows a specific project schedule based on task requirements. Each task must pass three states to be successfully completed upon arrival.
As Figure \ref{ADE} illustrates, as soon as the task is executed, it is added to the state of arrived tasks. Once the first agent registers for the task, a message (1) is released as a trigger and the task's state becomes \textit{“registered”}. When a task is in this state, the model provides a failure prediction ratio based on the agents’ reliability and experience levels. A message (2) released to report the \textit{registration phase failure}.
The registered task is associated with the duration following the triangular distribution reported in Part \ref{Task Completion2}. The registered task moves to a submitted state as soon as one of the registered agents makes a submission and releases a submission message. A submission message (3) worked as the second trigger and changed the task’s status to \textit{“submitted”}.
Once the task is moved to the submitted state, the failure prediction model switches to the submissions phase and continues the prediction based on the task submissions ratio in a time interval. In this moment, a new message (4) was sent out to report the \textit{submission phase failure}.
When the submissions deadline is over, the task moves to the peer review phase. Another message (5) is released to change the task’s status to from submitted to \textit{“peer review”}. If the quality score is greater or equal to 75, the task is reported as completed; otherwise, if the score is less than 75, the task is marked as failed and will be re-posted. Finally, the completed task is reported to the requester company and its sequential task will post in the system to be performed.

% \bigbreak
\subsection{	Micro-level: Agent Model}\label{alorithm}
   
Agent arrival in the AB model is an event that follows a Poisson distribution with $\gamma$ = 800. Once the model runs, agents start arriving in the model. When an agent decides to register for a task, the agent’s personal profile gets updated with new registered tasks. 
In this model, an agent’s registration decision follows an event with a rate of one registration per day. Also, none of the agents can work on more than five open tasks simultaneously. If the agent meets the above criteria and s/he has the required skillset to perform the task, s/he can decide to register for the task. In the registration decision phase, all agents that meet the above criteria are assigned to a random number in range of (0,1). If an agent’s random number is greater or equal to 0.8, s/he will register for the task. As soon as the agent registers for the task, the task’s status changes to registered. Moreover, according to the empirical findings, the average competition level per task is 18 \cite{yang2015award}. Therefore, in this model, the probability of an agent registering for a task with more than 18 assigned registrants follows Bernoulli distribution (P= 0.3), and the probability of an agent competing on a task with assigned registrants is less than or equal 18 is 1.
If an agent registers for a task, s/he can decide to make a submission for the same task. Agents’ submissions decisions for registered tasks follow an event with a rate of 0.51 submissions per day.
For deciding to make a submission, each agent is assigned to a random number in the range of (0,1). If the product of the random number and the agent’s registration probability, based on the agent experience belt, is less than 0.051, the agent can make a submission for the task. Once the agent makes a submission, the task’s state changes from registered to submitted.
After sending the submission message, the number of submissions updates in the agent’s profile and his or her reliability and rating factor is updated. The reliability of agents who make a submission follows a distribution with an average of 10\%. In this step, if the random score assigned to the submitted task is greater than 75, the agent will be reported as the winner. Algorithm \ref{alg:ABM} in appendix-\ref{finalalgorithm} presents the associated pseudo code with the AB model in the simulation.

\begin{figure*}[ht!]
\centering
\includegraphics[width=1\textwidth,  height=0.2\textwidth]{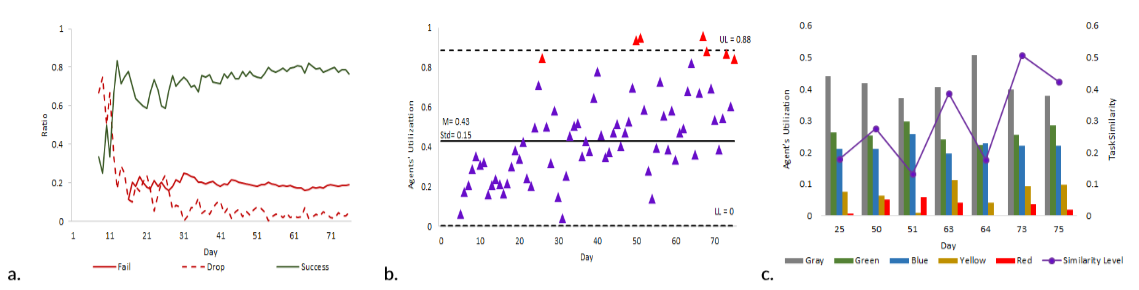}
\caption{Platform Performance: a. Simulation Platform Status,   b. Agent Utilization,   c. Agents’ Availability }
\label{performance}
\end{figure*}

\section{Evaluation}
To investigate the impact of task openness and agent diversity, the following research questions were formulated and studied in this paper:

\textit{RQ1(Tasks’ Openness)}: How much openness in the pool of open tasks can minimize the task failure ratio?
The reliability of qualified task submissions is measured based on the degree of task similarity in the pool of open tasks.

\textit{RQ2(Agent Diversity)}: How diverse should attracted agents be per task to receive a higher chance of qualified submissions? 
The consistency of agents’ availability in response to a task and the level of qualified performance is measured based on an agent’s experience level in the competition. %The resulting {\model} is illustrated in Figure \ref{overview}. This model addresses decision-making in terms of planning and understanding purposes \cite{gordon1961general}.

% \begin{figure}[ht!]
% \centering
% \includegraphics[width=1\columnwidth]{flow.png}
% \caption{Main Flow of Proposed Framework and Relationship to Research Questions}
% \label{flow}
% \end{figure}

\subsection{Performance of Failure Prediction Simulation in the Platform}

\subsubsection{\textbf{Platform Evaluation Design}}

The success of a CSD platform is based on the level of task failure in the platform. Since task failure is the result of agents’ performance in the platform, it is very important to understand agents’ utilization in the CSD platform. To address that end, the simulation was updated with the associated schedule of the example project and ran 30 times under the basic setting of simulation design. Each run lasted 60 days. During the simulation time, if a task failed due to a lack of qualified submissions or receiving zero submissions, it was reposted.
The last step was to understand the accuracy of failure prediction in a different task state. To do so, the mean relative error (${MRE}$) of each task failure prediction model (${FP_{zy}}$) based on the available actual failed task (${AF_{zy}}$) data gathered from TopCoder in the same day was calculated, where, ${z}$ is the time and ${y}$ is task state, as displayed below.

\[
{MRE_{zy}} = \frac{\sum_{z,y =0}^{l,2}{AF_{zy}}- \sum_{z,y=0}^{l,2}{FP_{zy}}} {\sum_{z,y =0}^{l,2}{AF_{zy}}}
\]
The t-test was also applied to the prediction results in each state to confirm the models’ accuracy.

\subsubsection{\textbf{Platform Performance}}

As it is presented in Figure \ref{performance}-a, the average success ratio in the platform while running the project was 71\% on average. The average none-qualified submissions ratio was 19\% and the average zero submission ratio was 7\%. Therefore, the average task failure ratio was 13\%. Also, the mean relative error (${MRE}$) of the failure prediction models was only 1.1\% in the registration phase and 2\% in the submissions phase. The results of the t-test on 60 observations per failure prediction model revealed that the probability of error in failure prediction in the registration phase was almost 0 and the error in the submissions phase was almost 1\% with zero hypothesized mean difference in both states. This result shows that the simulation is performing accurately. The Pearson correlation provides a incredibly close correlation between the actual failure for the project based on the data set and failure simulation results in the simulation. According to this test, the correlation in the registration phase was 0.21 and in the submission phase it was 0.42 with p-value less than 0.001.

% \begin{figure}[ht!]
% \centering
% \includegraphics[width=1\columnwidth]{SPstatus.png}
% \caption{Main Flow of Proposed Framework and Relationship to Research Questions}
% \label{SPstatus}
% \end{figure}

Additionally, it was reported that the agents’ utilization of the platform impacted the platform success ratio. After a deeper examination of the agents’ utilization of the platform at the time of sample project, as displayed in Figure \ref{performance}-b, it was concluded that the average agents’ utilization factor was 43\% with a standard deviation of 0.15. This information can be used to create a control chart to study the agents’ utilization of the platform. Figure 8 displays the control chart of agent utilization in the platform. Unlike an industrial control chart where the goal is to keep the quality as close as possible to the average, in CSD, the goal is to have utilization as high as possible and closer to the upper level line. Therefore, a closer examination occurred for the days that had a utilization factor near to or exceeding the upper level.
% \begin{figure}[ht!]
% \centering
% \includegraphics[width=0.9\columnwidth]{AgentUtilization.png}
% \caption{Main Flow of Proposed Framework and Relationship to Research Questions}
% \label{AgentUtilization}
% \end{figure}
Analyzing days with a higher level of agents’ utilization factor revealed that, on days with the highest utility factor, the average similarity of the open pool of task was less than 60\%. Interestingly enough, in those days, the availability of agents in the middle-ranking clusters (i.e. green and blue belts) were higher and relatively close to each other, as displayed in Figure \ref{performance}-c. 
% \begin{figure}[ht!]
% \centering
% \includegraphics[width=0.95\columnwidth]{AgentAvailability.png}
% \caption{Main Flow of Proposed Framework and Relationship to Research Questions}
% \label{AgentAvailability}
% \end{figure}
These results encouraged the team to design and test two different scenarios to answer the research questions that will be discussed in the following section.

\subsection{Evaluation Scenarios}
The aim of {\model}is to understand the optimal level of openness in the pool of available tasks and diversity in the competition level in order to minimize the task failure ratio in any CSD platform. In order to achieve the research goal, two different scenarios were proposed:

\textit{Scenario 1 (Task Openness)}: To successfully crowdsource a software project in a crowdsourcing platform, not only is it important to fully understand the project’s task dependencies, but also, it is vital to know the impact of available open tasks on each other within the platform. Different task similarity levels may represent task difficulty, task size, and task priority \cite{yang2015award}. Also, the degree of task similarity represents task openness in the pool of open tasks in the platform. Therefore, understanding workers’ behavior and performance based on the degree of task similarity in the pool of open tasks is very helpful to present more effective task planning.
Different researchers discuss that shorter task duration is associated with less complex work. This relationship may impact a worker’s decision-making processes in terms of taking tasks because he or she may want to achieve a higher number of tasks. Also, it is reported that there is a general negative correlation between a monetary prize as one of the elements of task similarity and worker behavior \cite{yang2015award}; workers are more attracted to tasks with a certain range of associated monetary prizes (i.e. \$750) \cite{yang2015award}. Therefore, task similarity can be an important factor in task failure rate.
Empirical studies demonstrate that, even though a higher similarity level among tasks leads to higher registration, it also results in lower submissions and a higher failure ratio. Scenario 1 aims to investigate the impact of task similarity level on agents’ performance in the CSD platform. In this scenario, four different testing policies based on previous empirical findings are tested. Each testing policy provides an arriving task with a limited similarity rate into the platform. The results of the four different testing policies were compared to each other.

\begin{figure*}[ht!]
\centering
\includegraphics[width=1\textwidth, height=0.3\textwidth]{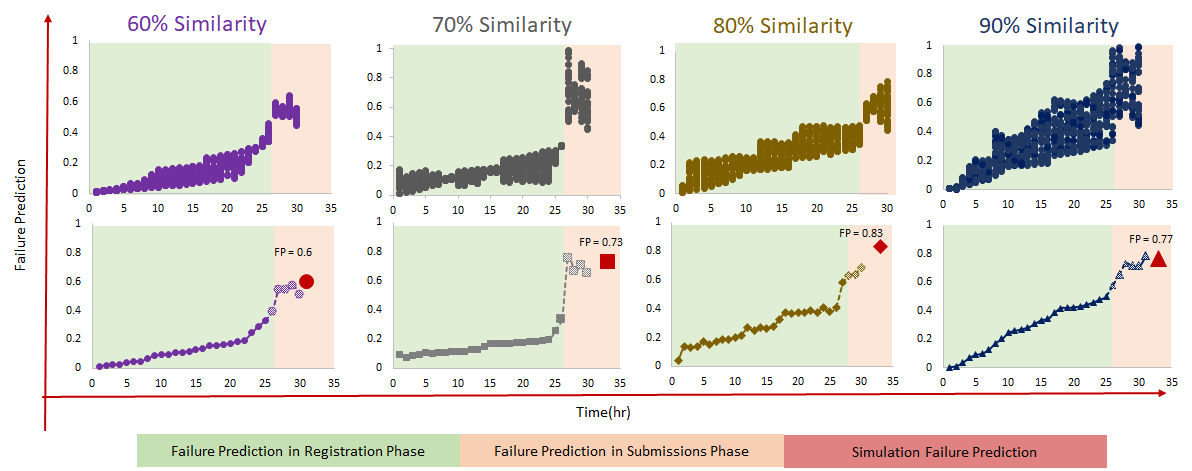}
\caption{Scenario 1, Impact of Task Similarity on Task Failure}
\label{Scenario1}
\end{figure*}
\textit{Scenario 2 (Agents’ Diversity)}: According to TopCoder, higher rating levels among agents represents a higher level of experience. Also, it is reported that a higher reliability factor among agents presents the highest chance of them submitting qualified tasks \cite{ye2015crowd}\cite{yu2015efficient}. In general, it is expected that more highly rated agents are associated with a higher reliability factor due to their experience. However, the reliability factor measure is based on the number of agents’ submissions and it is not related to a submission’s quality. It is reported that more highly experienced agents may take advantage of their reputation and apply cheap talk to the task \cite{archak2010money}. Scenario 2 aims to investigate the impact of diversity among agents in terms of agents’ experience level on agents’ performance.
In this scenario, four different testing policies were reviewed based on different agents’ experience levels as measured by TopCoder\footnote{ \url{https://www.Topcoder.com/member-onboarding/understanding-your-Topcoder-rating}}. Each testing policy attracted agents with a specific experience level to register for the chosen task, and agents’ behavior and performance level were reported. The results of the different testing policies were compared to each other.

\subsection{Results}
To apply evaluation scenarios, the motivation example was simulated. In the motivation example, task 8 was re-posted seven times until it was completed. This section presents simulation analysis and results under both scenarios.

\subsubsection{\textbf{Scenario 1, Task Openness}}

New arrival tasks in the platform joined the pool of open tasks and waited in the queue to be registered by agents. It is reported that agents are often more attracted to tasks with similar monetary prizes \cite{yang2015award}, task complexities, and context to each other \cite{difallah2016scheduling}\cite{khazankin2011qos}. This fact creates a demand market with an associated similarity factor to each task. Workers use this factor as one of the inputs to their decision-making process. Therefore, it is expected that a higher number of available similar tasks directly impacts the level of competition per task. The impact of this relationship was examined in this scenario.
According to empirical analysis, new arrival tasks compete with tasks with average similarity of 69\% per week. In Scenario 1, it was assumed that the project manager can control the openness of the pool of open tasks in terms of task similarity at the time of their task posting. The simulation ran 30 times per testing scenario and the average of all the runs was used in this scenario. 
% Four different testing policies based on degree of task similarity are discussed. The results of this study report the task failure prediction ratio in different states of a task. This scenario ran 120 times, more specifically 30 times under four different testing scenarios. Figure \ref{Scenario1} illustrates the failure prediction per testing policies.
% and Table \ref{S-Senario1} summarizes agents’ participation per testing policy in Scenario 1. 

% \begin{table}[!ht]
% \caption{Summary of Scenario 1} % title of Table
% \centering % used for centering table
% \begin{tabular}{p{0.4cm} p{2cm} p{0.7cm} p{0.7cm} p{0.7cm}p{0.7cm}}
% \hline
% \\
% \multicolumn{2}{c}{Scenario 1} & TP1 &  TP2 & TP3 & TP4\\ %[0.5ex] % inserts table
% \hline%\hline % inserts single horizontal line
% \\
% \parbox[t]{2mm}{\multirow{1}{*}{\rotatebox[origin=c]{90}{Task Status}}}
% & Fail & 18 & 22 & 25 & 23 \\
% \\
% & Success & 12 & 8 & 5 & 7 \\
% \\
% & Failure Prediction & 60\% & 73\% & 83\% & 77\% \\
% \hline
% \label{S-Senario1}
% \end{tabular}
% \end{table}

%\textit{Testing Policy 1: 60\% Degree of Task Similarity }

During the first testing policy, project manager posted the task to the pool of tasks with degree of similarity equal to 60\%. At the early registration stage the task received 1\% failure prediction, By the time that first submission happened, (the 26\textsuperscript{th}), failure prediction increased up to 35\%. At time 26, failure prediction shifted to the submission state and increased to 45\%. Unfortunately, the task only received four submissions out of on average 48 registrations, which increased the final failure prediction to 55\%. The results of the simulation indicated that the task failed for 18 out of 30 runs, which provides an actual failure ratio of 60\%. Interestingly during this policy task attracted diverse level of agents from all the available belts.

In the second testing policy, task was posted to the pool of open tasks with the the similarity level of 70\%. This time the task attracted 36 registrants and received only three submissions.  
The simulation under this try resulted in the task failing for 22 out of 30 runs i.e. 73\% task failure rate. Moreover, this task attracted competitors from all different ranking agents but yellow and only received submissions from  gray and green agents. 

During the third testing policy, task competed in a pool of tasks with an 80\% task similarity level. This policy the task only received on average one submission for an average of 23 registrations. Also, the task failed for 25 out of 30 runs, which provides an 83\% failure rate. During this test, the task attracted mostly gray and green belt agents, and some blue agents.  Also, tasks were only submitted by gray and green belt agents.

In the final testing policy, manager chose to post the task in the pool of 90\% task similarity. This policy received on average two submissions out of 32 registrations. The simulation resulted in 23 failed tasks out of 30 runs, representing a 77\% task failure rate. While increasing the task similarity level attracted all type of agents in terms of belt rankings, only the gray and green belt groups made submissions.
The details of each testing policy in scenario 2 is explained in appendix-\ref{Task Openness}.

% \bigbreak
% \textit{Discussion and Findings }
\textit{\textbf{Discussion and Findings.}}
It seems that the submissions ratio decreases when the level of task openness in the pool of open tasks decreases. In other words, the submission ratio falls when the degree of task similarity rises. Interestingly, tasks with a similarity level of 70\% and 60\% attracted more reliable agents in terms of submissions. Moreover, a higher similarity level among tasks resulted in a higher chance of task failure. This could be based on the fact that the group of agents attracted to the task have similar skillsets. This result confirms the fact that agents are more interested in lower levels of switching concepts. Also, a higher level of similarity among available tasks in the pool of open tasks may provide a higher chance of cheap talk \cite{archak2010money} per task due to a higher number of available options for active agents.
Further investigation revealed that the lowest and highest degrees of similarity ( i.e 60\% and 90\%) led to higher diversity in terms of agents’ rating levels, while middle-level degrees of task similarity attracted less diverse agents with lower level ratings. One reason for this grouping is that lower belt agents have a lower level of skillsets and experience, which makes them more interested in working on similar tasks \cite{difallah2016scheduling}.
Task 8 of the motivation example was tested with the findings of Scenario 1. Interestingly, the result showed thatif task 8 was posted with two days delay on April 4\textsuperscript{th}, the failure prediction would have dropped from 73\% to 60\% and shorten project schedule by 28 days. The impact of scenario 1 on task 8 is explained in details in appendix-\ref{SCE1Task8}.

% On April 2\textsuperscript{nd} , the day task 8 was posted in the platform, the average similarity among the pool of open tasks was 75\%, as displayed in Figure \ref{T8senario}. The average similarity of open tasks on April 1\textsuperscript{st}, April 5\textsuperscript{th}  and April 6\textsuperscript{th} was under 70\%, while on April 4\textsuperscript{th} it is under 60\%. 
% According to findings of Scenario 1, if task 8 was posted with two days delay on April 4\textsuperscript{th}, the failure prediction would have dropped from 73\% to 60\% and shorten project schedule by 28 days.
% Interestingly, when the task was successfully completed as task 18, the average task similarity in the pool of open tasks was 54\%.

% \begin{figure}[ht!]
% \centering
% \includegraphics[width=0.95\columnwidth]{T8senario1.png}
% \caption{Impact of Scenario 1 on Task 8}
% \label{T8senario}
% \end{figure}

\textbf{\underline{Finding}}: A higher level of openness (60\%) in the pool of tasks leads to a lower chance of task failure for the crowdsourced project.

% \bigbreak
\subsubsection{\textbf{Scenario 2, Agents’ Diversity}}

%Reliability in receiving a submission for a task by a registered agent is one of the main factors affecting a task’s success. The reliability factor, based on the number of an agent’s submissions, is not related to the quality of the task submission. However, it is expected that higher-rated agents have higher reliability in making a submission.

\begin{figure*}[ht!]
\centering
\includegraphics[width=1\textwidth, height=0.38\textwidth]{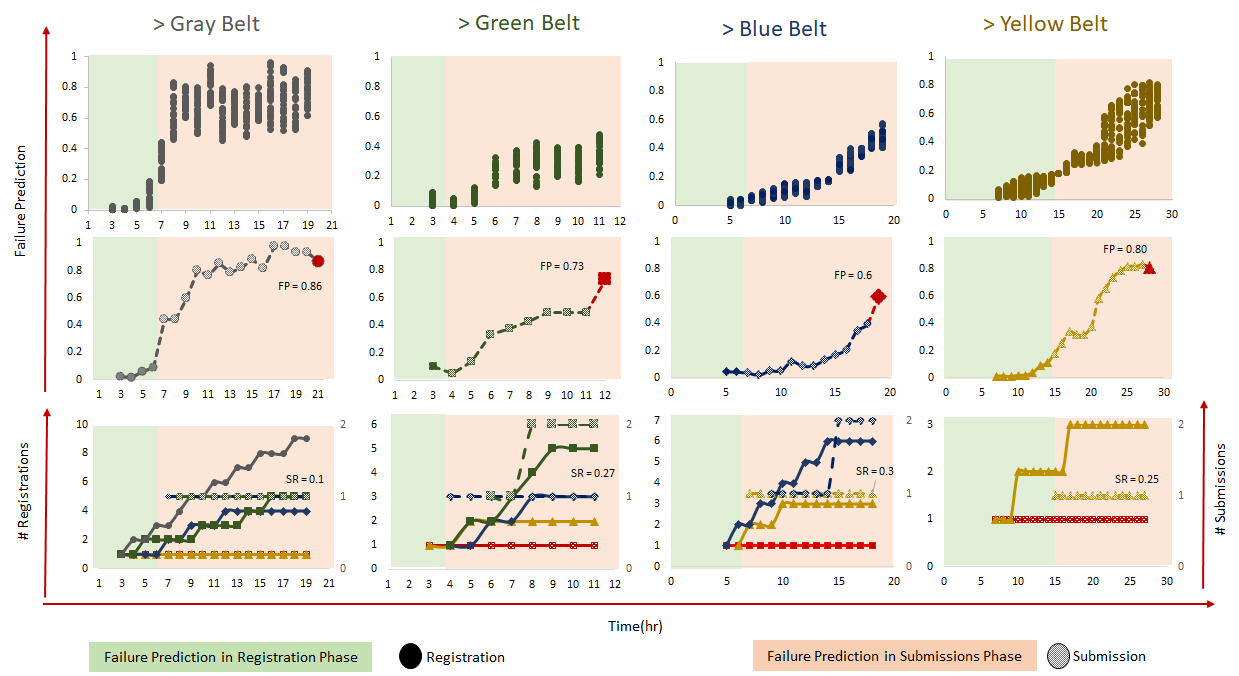}
\caption{Scenario 2, Impact of Agents’ Experience on Task Failure}
\label{Scenario2}
\end{figure*}
Generally, agents do attempt to register for a higher number of tasks than they can complete \cite{mao2017survey}, but even so, higher-rated agents may apply the cheap talk strategy to ease the competition level and guarantee their win \cite{archak2010money}. In this scenario, the impact of the task’s restricting access availability was analyzed for agents with certain belt ratings to address that end. The simulation ran under four different availability policies and each policy ran 30 times. Figure \ref{Scenario2} presents the failure prediction and submissions ratios under the different testing policies.
%and Table \ref{S-Senario2} summarizes agents’ participation in the different testing policies in Scenario 2.

% \begin{table}[!ht]
% \caption{Summary of Scenario 2} % title of Table
% \centering % used for centering table
% \begin{tabular}{p{0.4cm} p{1cm} p{1cm} p{0.7cm} p{0.7cm}p{0.7cm}p{0.7cm}}
% \hline
% \\
%  \multicolumn{3}{c}{Scenario 2} & TP1 &  TP2 & TP3 & TP4\\ %[0.5ex] % inserts table
% \hline%\hline % inserts single horizontal line
% \\
% \parbox[t]{2mm}{\multirow{1}{*}{\rotatebox[origin=c]{90}{Task Status}}}
% &\multicolumn{2}{c}{Fail} & 18 & 22 & 25 & 23 \\
% \\
% &\multicolumn{2}{c}{Success} & 12 & 8 & 5 & 7 \\
% \\
% &\multicolumn{2}{c}{Failure Prediction} & 60\% & 73\% & 83\% & 77\% \\
% \hline
% \\
% \parbox[t]{2mm}{\multirow{18}{*}{\rotatebox[origin=c]{90}{Agents' Participation}}}
% &\multirow{3}{*}{Gray} 
% & Reg &  &  &  & 44\% \\
% \\
% && sub & &  &  & 18\%  \\
% \\
% &\multirow{3}{*}{Green} 
% & Reg &  &  &  53\%& 29\% \\
% \\
% && sub &  &  & 50\% & 45\%  \\
% \\
% &\multirow{3}{*}{Blue} 
% & Reg &  & 62\% &  26\% & 17\% \\
% \\
% && sub &  &  67\% & 50\% & 36\%  \\
% \\
% &\multirow{3}{*}{Yellow} 
% & Reg &  82\% & 31\% &  16\% & 7\% \\
% \\
% && sub &  80\% & 33\% & 0\% & 0\%  \\
% \\
% &\multirow{3}{*}{Red} 
% & Reg &  18\% & 8\% &  5\% & 2\% \\
% \\
% && sub &  20\% & 0\% & 0\% & 0\%  \\
% \\
% \hline
% \label{S-Senario2}
% \end{tabular}
% \end{table}

% \bigbreak
% \textit{Testing Policy 1: Red and Yellow}

Under first testing policy, the project manager provide access to the agents with a belt rating of more than 1,500 points to compete on the task(i.e. yellow and red belt). The first registration happened at time seven and had a failure prediction ratio around 1\%; and at time 15 the first submission occurred and increased the failure prediction ratio to 18\%. The task submissions deadline continued until time 27, and the failure prediction increased to 83\%. On average, four agents were interested in this task and only one agent made a submission i.e. 80\% unqualified submission. Yellow belt agents made up 82\% of participants and 80\% of submissions..

% \bigbreak
% \textit{Testing Policy 2: Red, Yellow and Blue}

During second testing policy, all agents with a rating above 1200 points (Red, Yellow and Blue belt) were permitted to compete on the task.  On average, the task received ten registrations and three submissions under this policy, which provides a 60\% chance of task failure. Of the attracted agents,majority of registration and submission were blue belts agents, while red agents did not make any submission. .

% \bigbreak
% \textit{Testing Policy 3: All but Gray}

In the third testing policy, the project manager opened the task to all available agents with 900 rating points or above (all but gray belt). The higher diversity of belt levels attracted 12 registrants and a maximum of four submissions. There was a 73\% chance that the task received unqualified submissions.
Under this policy, while all different rating belt registered for task, against the expectation, more higher ranked agents did not make any submissions.Blue and green belt agents made equal level of submissions.

% \bigbreak
% \textit{Testing Policy 4: All Welcome}

In the last testing policy, all available agents in the platform were welcome to compete on the posted task. Simulation results revealed that on average, the task received 20 registrations and only one submission. The chance of receiving an unqualified submission under this policy was approximately 86\%.
This policy welcomed a wide diversity of available agents in the platform as competitors on this task. Interestingly, while gray belt agents were the highest level of participants, they made very little of the submissions. Green belt agents made most of submissions, following by blue belt agents. Yellow and red belt agents did not submit for the tasks under this policy at all.
The details of each testing policy in Scenario 2 is explained in appendix-\ref{Agents’ Diversity}.

% \bigbreak

\textit{\textbf{Discussion and Finding.}}
It was expected that a higher level of diversity in agents’ rating and experience provided lowered task failures. However, the simulation results reported that a higher level of diversity among agents led to the highest level of failure prediction per task. Interestingly, the lowest level of diversity led to the same result, while attracting mid-level agents provided the lowest level of task failure prediction. The result of policies three and four in Scenario 2, not only provided higher task submission ratio of 30\% and 27\%, but they also resulted in lower failure prediction rates of 60\% and 73\% respectively. This result supports our empirical studies’ reports that middle-rated agents are more reliable in making a submission \cite{saremi2017leveraging}.
The findings of Scenario 2 were tested on task 8 from the motivation example. The result of the test indicated that decreased from 86\% to 73\% and the project schedule would have been 26 days shorter. The impact of scenario 2 on task 8 is explained in details in appendix-\ref{SCE2Task8}.

% As displayed in Figure \ref{T8senario2}, the day that task 8 was posted in the platform, the availability of middle level agents (blue and green belts) was less than 5\%. The availability of mid-level agents was considerably higher on April 1\textsuperscript{st}, April 3\textsuperscript{rd}, April 4\textsuperscript{th}, and 6\textsuperscript{th}. Interestingly, according to empirical data on April 3\textsuperscript{rd} and April 4\textsuperscript{th}, the platform hosted the highest level of mid-level agents. It seems that if the task had been posted on either of those days, it might have had a higher chance of success. 
%  According to the result of Scenario 2, if the task had been posted on April 6\textsuperscript{th}, the chances of failure would have decreased from 86\% to 73\% and the project schedule would have been 26 days shorter. Interestingly, based on available data, when the task was successfully completed as task 18, the average of mid-level agents’ availability was 7\%.

% \begin{figure}[ht!]
% \centering
% \includegraphics[width=0.9\columnwidth]{T8senario2.png}
% \caption{Impact of Scenario 2 on Task 8}
% \label{T8senario2}
% \end{figure}

\textbf{\underline{Finding}}: Attracting mid-level experienced agents(blue and green belts) to the task competition decreases the chance of task failure.

% \bigbreak
\subsubsection{\textbf{Limitations}}
First, the study only focuses on competitive CSD platforms. Many more platforms do exist, and even though the results achieved are based on a comprehensive set of about 5,000 development tasks, the results cannot be claimed as externally valid. There is no guarantee that the same results would remain the same in other types of CSD platforms.
Second, there are many different factors that may influence task similarity and workers’ utility decision in task selection and completion. {\model}is designed based on known task attributes in TopCoder. Different similarity algorithms may lead us to different but almost similar results. 
Third, the result is only based on task similarity and worker's reliability. The worker's network and communication were not considered in this research. In future we need to add this level of research to the existing one.

\section{Related Work}

\subsection{Platform Workflow in CSD}
% A successful crowdsourcing platform contains three determinants: the characteristics of the project; the composition of the crowd; and the relationship among key players \cite{mao2017survey}. 
% A systematic development process in such platform a starts from a requirements phase, where the project goals, task plan, and budget estimation are recognized. This will be performed through communication between the project manager, who may come from the crowd or the platform, and the requester, who pays for the solutions offered by the crowd. The outcome will be a set of requirements and specifications. These requirements are used as the input for the future architecture phase, while the application is decomposed into multiple components \cite{mingozzi1998exact}.

In CSD workflow , the requester company divides the project into many small tasks, prepares task descriptions, and distributes tasks through the platform. Each task is tagged with a pre-specified prize as the award \cite{yang2015award}\cite{mao2013pricing} to winners and a required schedule deadline to complete. On average, most of the tasks have a life span of 2-4 weeks from the first day of registration. 
Crowd software workers browse and register to work on selected tasks, and then submit the work products once completed. After workers submit the final submissions, the files will be evaluated by experts and experienced workers, through a peer review process, to check the code quality and/or document quality \cite{mao2017survey}. The number of submissions and the associated evaluated scores replicate the level of success in task success. In TopCoder, usually the award goes to the top 1 or 2 winners with the highest scores. If there are zero submissions, or no qualified submissions, the task will be treated as starved or cancelled.

\subsection{Simulation Methods}
Simulation techniques have been demonstrated to be a powerful method in modeling and understanding NP-hard problems in software engineering processes, such as system dynamics as well as discrete event and agent-based simulation models \cite{zhang2011impact}. 
Discrete Event Simulation (DES) Methods are able to create a step by step simulation of the flow of the entities and their impact in the system. Any event happening will lead to a change in the state of the system. All the entities, attributes, events and relationships among stakeholders should be defined in the model of the system \cite{greening2016agile}. The main advantage of DES is capturing the actual process level details and representing each work product as a unique individual in the development process \cite{gao2015constructing}\cite{law2000simulation}.
Agent based simulation (ABS) techniques have been employed to simulate individuals and social groups’ behaviors in crowdsourcing, as well as the interaction between requestors and software workers \cite{zhang2012toward}\cite{si2014encoding}\cite{kaufmann2011more}. ABS helps to present more details of development process. While in ABS, the number of stakeholders (agents) is constant. In a systems dynamic (SD) simulation the number of stakeholders will vary by passing time continuously. Such difference makes it possible to predict the number of active software workers and their arrival distribution \cite{turner2012modeling}\cite{pfahl2000using}. 
In addition, systems dynamic simulation methods provide the possibility of changing one or several factors (attributes) while the remaining ones are unchanged, thus supporting all possible scenarios to be checked in order to make decisions based on managerial policies \cite{gao2015constructing}\cite{setamanit2006planning}. Also, it is difficult to identify a complex sequence of activities in the process in SD models \cite{setamanit2006planning}.

In practice, a mixture simulation model merging systems dynamic and discrete event models has been used to support project planning and improvement of global software development \cite{pfahl2000using}. Also, a hybrid simulation model based on a combination of systems dynamic, discrete event and agent-based simulation techniques was presented for a Kanban scheduling system \cite{turner2012modeling}\cite{madachy2007software}. Nevertheless, crowdsourced platforms have been simulated via systems dynamic modeling \cite{saremi2015dynamic}. 
It is of our interest to study the hybrid simulation approach in CSD, in order to expand and improve understanding of crowdsourcing decision making. More specifically, impact of three different component in crowdsourcing task scheduling.

% \subsection{Task Scheduling in Software Engineering}
% Task scheduling in software engineering is a sequence of time dependent functions/tasks that make a project \cite{tregubov2015simulation}. Generally, scheduling is subject to minimizing cost and time \cite{tajedin2013determinants}, while maximizing quality, resource allocation and task priority \cite{alba2006efficient}. 
% Dependencies among tasks and resources makes task scheduling an NP-hard problem. In the traditional software development process, task scheduling is the next step after defining requirement development and deciding on task decomposition \cite{amiri2015new}. Since more than half of software requirements are interdependent and only a few of them are independent, scheduling is complex \cite{carlshamre2001industrial}. 

% Traditional scheduling techniques are not considering human resource allocation, or if they do, they are not covering resources with various skills.  Since humans are the main resource in software projects, scheduling can be more flexible compared to other industries \cite{kittur2008crowdsourcing}. To overcome this issue, project managers are using global software development techniques to reduce cost and shorten release time. Although, different time zone and resource location may add to the scheduling challenges \cite{sorokin2008utility}. CSD is one of the methods that help reducing software development challenges.

\subsection{Task Scheduling in CSD}
Batching tasks is an effective scheduling method in CSD to reduce the complexity of tasks and will dramatically reduce cost \cite{marcus2011human}.  Batching crowdsourcing tasks would lead to a faster result than approaches, which keeps workers separate and is also faster than the average of the fastest individual worker \cite{kittur2013future}. There is a theoretical minimum batch size for every project as one of the principles of product development flow \cite{reinertsen2009celeritas} To some extent, the success of software crowdsourcing is associated with reduced batch size in small tasks. Tasks can be batched based on their similarity in skill requirements, associate price, priority and development phase etc.
Another useful scheduling method in CSD is parallelism. This method will create the chance to utilize a greater pool of workers \cite{ruhe2005art}, help workers to specialize and complete the task in less time, drive solutions, help the requester to clearly understand how workers decide to compete on a task, and analyze the crowd workers' performance \cite{faradani2011s}. Shorter schedule planning can be one of the biggest advantages of using parallelism for managers \cite{archak2010money}.

% Moreover, complex projects cannot be performed based on simple available parallel approaches. Complex projects have more dependencies and multiple changing requirements \cite{kittur2013future}, that requires different workers with different levels of expertise. Therefore, this is one of the main challenges in applying an effective method to schedule decomposed projects \cite{latoza2013crowd}. Considering the fact that coordinating workers is difficult among a distributed global crowd, in most cases, organizational coordination techniques can be applied to crowd work as well \cite{marchandh}, such as programming and feedback as general coordination methods \cite{ross2010crowdworkers}.  

%\subsection{Available Scheduling Methods in CSD}
Delay scheduling \cite{tajedin2013determinants} is the first presented scheduling method for crowdsourced projects to maximize the probability of a worker receiving tasks from the same batch of tasks they were working on. Extension of this idea introduced a new method called fair sharing schedule \cite{alba2006efficient}. In this method, heterogeneous resources would be shared among all tasks with different demands, which ensures that all tasks would receive the same amount of resources to be fair. 
% For example, this method was used in Hadoop Yarn.
Later on, weighted fair sharing (WFS) \cite{barreto2008staffing} was presented as a method to schedule batches based on their priority. Tasks with higher priority are introduced first. Another proposed crowd scheduling method is quality of submissions (QOS) \cite{khazankin2011qos}, a skill-based scheduling method with the purpose of minimizing scheduling while maximizing quality by assigning the task to the most available qualified worker. This method was created by extending standards of Web Service Level Agreement (WSLA) \cite{regnell2005market}. The third available method is game with a purpose \cite{cusumano2004business}, in which a task will not be started unless a certain number of workers registered for it.
Similarly, CrowdControl \cite{rajan2013crowdcontrol}, provides a scheduling approach that assigns
tasks to workers not only based on their historical performance but also their learning curve of performing tasks. SmartCrowd \cite{roy2014optimization} schedules tasks based on worker skills and their reward requirements to perform the task. Using above scheduling methods, HIT-Bundle (Human Intelligent Task) \cite{difallah2016scheduling} provides a batch container which schedules heterogeneous tasks into the platform from different batches. This method makes a more successful outcome by applying different scheduling strategies at the same time. The most recent method is helping crowdsourcing-based service providers to meet completion time SLAs \cite{hirth2019task}. The system works based on the oldest task waiting time and runs a stimulative evaluation to recommend best scheduling strategy in order to reduce the task failure ratio.

\section{Conclusion}
To ensure an effective and successful crowdsourcing platform in such dynamic market, there is a need for improved understanding and visibility into the characteristics and risks associated with CSD processes. While most of the CSD approaches are based on the static aspects of CSD marketplaces, its dynamic aspects have often been ignored.  In this paper, we presented a hybrid simulation model {\model} to forecast crowdsourcing task failure risk in crowdsourcing platforms. {\model} is composed of three components: the macro-level systems dynamic simulation reflecting the crowdsourcing platform dynamics, the meso-level of discrete event simulation representing the task life cycle, and the micro-level agent-based simulation illustrating the crowd workers’ decision-making processes. {\model} is evaluated through three CSD decision scenarios to demonstrate its effectiveness, using a real-world historical dataset. The results show that, task openness in the platform in terms of task similarity and agents’ experience are two of the most effective factors that should be considered while planning for a more effective task execution. In details: 1) A pool of open tasks with a higher degree of task similarity in the platform leads to higher level of failure prediction, 2) Attracting agents with mid-level experience to compete on the task helps to achieve a lower level of failure prediction, 3) Agents with mid-level experience provide a higher level of task submissions ratio. {\model} empowers crowdsourcing managers to explore potential crowdsourcing outcomes with respect to different task scheduling options. In our ongoing and future work, we are considering providing a model to predict agents’ trust-ability besides agents’ reliability.

\begin{small}
\bibliography{mybibfile.bib}
\end{small}

\clearpage
%###########################################################

\setcounter{table}{0} \renewcommand{\thetable}{\arabic{table}.Appendix}
\setcounter{figure}{0} \renewcommand{\thefigure}{\arabic{figure}.Appendix}

\appendix

%%%%%%%%%%%%%%%%%%%%%%%%%%%%%%%%%%%%%%%%%%%%%%%%%%%%%%%%%%%%%%%%%%%%%%%%%%%%%%%%%%%%%%%%%%%

\subsection{Competition Model}

Systems Dynamic Simulation(SDS) at the top level of presented model, represent the overall relationship among task arrival, worker reliability, and task competition outcome, This model contains 14 variables including task, agent decision, workers’ performance, task similarity, worker profile, worker skillset, and different available crowdsourced markets. The relation among these factors provides different causal loops between different levels of the platform.
Table \ref{VSD} summarizes the structured attributes used in the systems dynamic simulation.

\begin{table}[!ht]
\caption{Variable Used in Systems Dynamic Model} % title of Table
\centering % used for centering table
%\begin{tabular}{p{2cm}p{4cm}p{1.5cm}}
\begin{tabular}{p{2cm}p{4cm}p{1.5cm}}

\hline
Metric & Definition & Model Variables\\
\hline
\\
Task Similarity (${TSim}$) & Tuple of monitory prize, task duration, task complexity, required technology and task type. Range [0,1]& tTSim(double) \\
\\
Task Duration (${D}$) & Total available time from task registration start date to submission deadline. Range:(0, $\infty$) & duration(time)\\
\\
Failure Ratio (${FR}$) & Ratio of cancelled tasks per total tasks per platform. Range [0,1] & tFR(double) \\
\\
Task Arrival (${TAr}$) & Task uploaded to the platform to be worked on. Range: (0, $\infty$) & tTAr( int) \\
\\
Task Failure (${TF}$) & Task that does not receiving any qualified submissions. & tTF(int)\\
\\
Task Submissions Ratio (${TSR}$) & Percentage of submitted tasks to the total number of registered tasks in any given time. Range [0,1] & tTSR(int) \\
\\
Task Completion Ratio (${TCR}$) & Percentage of number of tasks that can pass the peer review to total number of registered tasks in any given time. Range [0,1] & tTCR(double) \\
\\
Task Drop Ratio (${TDR}$) & Percentage of number of tasks that is not submitted while is registered to total number of registered tasks in any given time. Range [0,1] & tTDR(double) \\
\\
Agent Decision (${AD}$) & Agent decides to register for a task and/or make a submission for the same task. & tAD(Register, Submit)\\
\\
Agents Performance (${AP}$) & Number of qualified submissions that an agent makes Range: (0, i) & tAP(double) \\
\\
Agent Skill Set (${ASS}$) & Workers’ list of skillsets in their profile. & tASS(Java, C)\\
\hline
\label{VSD}
\end{tabular}
\end{table}

%%%%%%%%%%%%%%%%%%%%%%%%%%%%%%%%%%%%%%%%%%%%%%%%%%%%%%%%%%%%%%%%%%%%%%%%%%%%%

% \subsection{Meso-level Task Completion: Discrete Event}

% Table \ref{VDE} summarizes the structured attributes used in the Discrete Event simulation.

\begin{table}[!ht]
\caption{Variable used in Discrete Event Simulation Model} % title of Table
\centering % used for centering table
\begin{tabular}{p{2cm}p{4cm}p{1.5cm}}

\hline
Metric & Definition & Model Variables\\
\hline
\\
Registered (${R}$) & Number of registrants that are willing to compete on total number of tasks during a specific period of time. Range: (0, r) & tregister(int) \\
\\
Task Duration (${D}$) & Total available time from task registration start date to submission deadline. Range:(0, $\infty$) & duration(time)\\
\\
Submitted (${S}$) & Number of submissions that a task receives by its submission deadline in specific period of time. Range: (0, \# registrants) & tsubmit(int) \\
\\
Peer Reviewed (${PR}$) & Process of reviewing a submitted task to check the quality of submissions. Range: (0, \# registrants) & tpeer(int) \\
\\
Completed (${C}$) & Qualified task that has successfully passed the peer review. & tcomp(int)\\
\\
Re-worked ((${RW}$) & Qualified task that needs some adjustment in order to pass the peer review. & trework(int) \\
\\
Failed (${F}$) & Non-qualified task that has not passed peer review. & tfail(int) \\
\\
State & Each task has a state that declares the task situation during the submission process. & state (Arrived, Registered, Submitted, Reviewed, Complete) \\
\hline
\label{VDE}
\end{tabular}
\end{table}

\subsection{Task Completion Model}

Task lifecycle, and the level of task complexity and priority in this paper is modeled by Discrete Event Simulation (DES).
Table \ref{VDE} summarizes the attributes used in the Discrete Event simulation.

%%%%%%%%%%%%%%%%%%%%%%%%%%%%%%%%%%%%%%%%%%%%%%%%%%%%%%%%%%%%%%%%%
\subsection{Agent Model}\label{finalalgorithm}

In order to simulating  various characteristics, decision-making processes, and performance of crowd workers, this model used agent based modeling. Crowd-workers are represented as agents in this model. 
Table \ref{VAB} summarizes the structured The attributes used in the Agent Based Simulation simulation, and Algorithm \ref{alg:ABM} presents the associated pseudo code with the agent based model in the simulation.

\begin{table}[!ht]
\caption{Variable Used in Agent-Based Simulation Model} % title of Table
\centering % used for centering table
\begin{tabular}{p{1.5cm}p{3cm}p{2cm}}

\hline
Metric & Definition & Model Variables\\
\hline
\\
Registering (${RA}$) & Number of tasks a worker registered for in specific period of time. Range: (0, i) & a\_registering (ArrayList${<Task>}$) \\
\\
Submitting (${SA}$) & Number of tasks a worker submits in specific period of time. Range: (0, \#registrants] & a\_submitting (ArrayList${<Task>}$)\\
\\
Quality (${Q}$) & The quality of submitted task based on associated score. Range: (0, 1) & tQ(int) \\
\\
Reliability (${Re}$) & The percentage of successful task submissions in a worker’s most recent 15 task registrations. Range: (0, 1) & tRe(int) \\
\\
Trust-ability (${TA}$) & Percentage of number of qualified submissions among total number of submissions that a worker makes & tT-A(int)\\
\hline
\label{VAB}
\end{tabular}
\end{table}

\subsection{Scenario 1, Task Openness}\label{Task Openness}
 The details of four different testing policies based on scenario 1 are discussed as below:

\subsubsection{Testing Policy 1: 60\% Degree of Task Similarity }

The project manager posted a task in the pool with degree of similarity equal to 60\%. The result of this 
policy led to, on average, only 1\% failure prediction at the early registration stage; by the 26th time that the first submission happened, failure prediction increased up to 35\%. This fact demonstrates that, by passing time, the chance of receiving highly reliable agents to compete on a task is low; however, the average failure ratio in this state is 8\%. At time 26, failure prediction shifted to the submission state and increased to 45\%. Unfortunately, the task only received four submissions out of on average 48 registrations, which increased the failure prediction to 55\% on the task submission deadline. The results of the simulation indicated that the task failed for 18 out of 30 runs, which provides an actual failure ratio of 60\%. 

Also, applying this policy attracted a wide diversity of agents. The results revealed that competing for tasks in a pool of other tasks with 60\% similarity attracted on average 44\% of all gray belt agents, 22\% of all green belt agents, 21\% or all blue belt agents, 11\% of all yellow belt agents, and almost 6\% of all red belt agents.

\subsubsection{Testing Policy 2: 70\% Degree of Task Similarity  }

Under testing policy 2, the project manager kept the similarity level of the pool of open tasks at 70\%. At the first registration, the failure predictor predicted a 9\% task failure. With increased time passed and higher competition levels, the task failure prediction increased to 33\% in the registration phase. The average failure prediction in the registration state was 14\%. At time 26, the first submission was made, and the failure prediction increased to 75\%. Finally, at time 30, the task received its third and final submission with a failure prediction of 67\%. This task attracted 36 registrants and received only three submissions.  
The simulation under this policy resulted in the task failing for 22 out of 30 runs. This result indicates a 73\% task failure rate. Moreover, this task attracted competitors from the gray, green, blue and red ranking belts with a participation ratio of 37.5\%, 25\%, 25\% and 13\% respectively. Among all the participants, only gray and green agents made submissions. It seems that, under this policy, the task cannot attract high-reliability agents to make a submission; however, it successfully attracted more highly ranked agents to compete. 

\begin{algorithm*}[h]
\caption{Agent Decision Making Procedure}
    \label{alg:ABM}
\SetKwInOut{Input}{input}
\SetKwInOut{Result}{result}
\SetKwProg{ReturnSolution}{return}{}{}
\DontPrintSemicolon
\ReturnSolution{(Agent Registering Status$,$ 
         Agent Submitting Status$,$ 
         Agent Winning Status)}{
\Input{An agent from the pool of crowd agents,
A task from the pool of crowdsourced tasks}
\Result{Agent decision making process for: registering for a task, making a submission and passing peer review.}
\BlankLine
\nl \textbf{Agent registering for a task}
\BlankLine
\nl \textit{Create Arrival Event (1, day)}
\BlankLine
\nl \For{Task}{
\nl    \uIf{((Agent is Registering),(Agent Open List $<$ 5),(Agent Expertise = Task Requirements),(Agent Rating $>$ 0))}{
\nl        \If{(\# Registrants for Task $<$ 18 )}{
\nl         \If{(uniform distribution $>$ 0.80)}{
\nl         send \textit{\{"Agent Registered"\}}\;
\nl    		Task added to Agent Open List\;
\nl    		Agent added to Task Registrants List\;
\nl    		Task Rating = Agent Rating;}
    		}}
 \nl       \Else{
 \nl            \If{(Bernoulli(0.3) = 1)}{
 \nl            send \textit{\{"Agent Registered"\}}\;
 \nl			Task added to Agent Open List\;
 \nl \nl			Agent added to Task Registrants List\;
 \nl			Task Rating = Agent Rating\;
		      }}
		      }
\BlankLine
\nl \textbf{Agent submitting for a task}
\BlankLine
\nl \textit{Create Arrival Event (0.51, day)}
\BlankLine
 \nl \For{Task in Agent Open List}{
 \nl   \uIf{((Agent is Submitting),(Agent Open List $>$ 1))}{
  \nl       Generate a Random Number\;
 \nl       P(qualified Sub) Based on Table I\;
 \nl       \If{(P(qualified Sub) * Random) $<$ 0.051}{
\nl		    Send \textit{\{"Agent Submitted"}\}\;
\nl		    Agent Score Reported\;
\nl		    Task added to Agent Submitted List\;
\nl		    Agent added to Task Submitter List\;
    }
        }
 \nl  \For{Task in Agent Submitting List}{
 \nl       \uIf{Agent Score $>$ Task Score}{
\nl            Task Score = Agent  Score\;
\nl 		    Agent reported as Task Winner\;
\nl		    Task added to Agent Winner List\;
\nl		    Send \textit{\{"Agent Won"\}}\; 
			}

    }
}

}
\end{algorithm*}

%%%%%%%%%%%%%%%%%%%%%%%%%%%%%%%%%%%%

\subsubsection{Testing Policy 3: 80\% Degree of Task Similarity}

 Under this policy, the project manager created a pool of tasks with an 80\% task similarity level. However, at time one, the task provided a competition environment with only 1\% failure prediction; at time two, failure prediction rose to 22\%. This policy attracted slow-arriving agents to the competition. Only six agents arrived in the first 26 days before the task received its first submissions. At time 12, the failure ratio increased to 36\% from 26\%, and during time 16, it rose to 41\%. By the 18th run, the rate was at 47\%.  At time 26, the first submission occurred and pushed the failure prediction metrics to the submissions state. This task only received on average one submission for an average of 23 registrations. This low submission number increased the failure prediction to 63\% at time 26, and finally, to a failure prediction rate of 78\% on the 30th and final run. Also, the task failed for 25 out of 30 runs, which provides an 83\% failure rate.
Interestingly applying this policy resulted in a lower level of diversity among participants in terms of ranking belt. During this test, the task attracted on average, 50\% of gray belt agents, 45\% of green belt agents, and only 5\% of blue agents.  Also, tasks were only submitted by gray and green belt agents.

\subsubsection{Testing Policy 4: 90\% Degree of Task Similarity }

When the project manager followed policy four, the task faced the challenge of very slow agent attraction. The failure prediction rate started at almost 0\% and increased to 52\% in day 24. The task faced two jumps in attracting agents. The first jump occurred on day 4, when the failure prediction rate rose to 25\% from 14\%. The second jump happened on day 16, when the failure prediction reached 43\% due to attracting low reliable workers. This task received on average two submissions out of 32 registrations. The first submission happened on day 25 and the second one occurred on day 28. The task failure prediction in the submission state increased to an average of 78\% on day 30. The simulation resulted in 23 failed tasks out of 30 runs, representing a 77\% task failure rate.
While applying a higher level of task similarity did not help with lower failure predictions, the increased task similarity attracted a better diversity of agents in terms of belt rankings. Simulation results indicated that 45\% of participants were gray belt agents, 27\% green belt agents, 9\% blue belt agents, and 18\% yellow belt agents. However, only the gray and green belt groups made submissions, each for 50\%.

Table \ref{S-Senario1} summarizes agents’ participation per testing policy in Scenario 1. 

\begin{table}[!ht]
\caption{Summary of Scenario 1} % title of Table
\centering % used for centering table
\begin{tabular}{p{0.4cm} p{2cm} p{0.7cm} p{0.7cm} p{0.7cm}p{0.7cm}}
\hline
\\
\multicolumn{2}{c}{Scenario 1} & TP1 &  TP2 & TP3 & TP4\\ %[0.5ex] % inserts table
\hline%\hline % inserts single horizontal line
\\
\parbox[t]{2mm}{\multirow{1}{*}{\rotatebox[origin=c]{90}{Task Status}}}
& Fail & 18 & 22 & 25 & 23 \\
\\
& Success & 12 & 8 & 5 & 7 \\
\\
& Failure Prediction & 60\% & 73\% & 83\% & 77\% \\
\hline
\label{S-Senario1}
\end{tabular}
\end{table}

\subsubsection{Impact of Scenario 1 on Task 8} \label{SCE1Task8}

\begin{figure}[ht!]
\centering
\includegraphics[width=0.95\columnwidth]{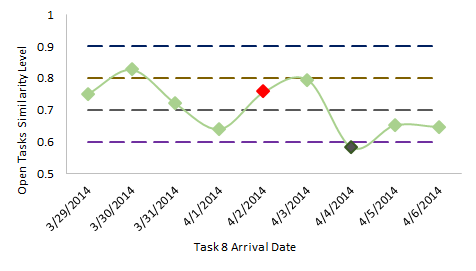}
\caption{Impact of Scenario 1 on Task 8}
\label{T8senario}
\end{figure}

On April 2\textsuperscript{nd} , the day task 8 was posted in the platform, the average similarity among the pool of open tasks was 75\%, as displayed in Figure \ref{T8senario}. The average similarity of open tasks on April 1\textsuperscript{st}, April 5\textsuperscript{th}  and April 6\textsuperscript{th} was under 70\%, while on April 4\textsuperscript{th} it is under 60\%. 
According to findings of Scenario 1, if task 8 was posted with two days delay on April 4\textsuperscript{th}, the failure prediction would have dropped from 73\% to 60\% and shorten project schedule by 28 days.
Interestingly, when the task was successfully completed as task 18, the average task similarity in the pool of open tasks was 54\%.

%%%%%%%%%%%%%%%%%%%%%%%%%%%%%%%

\subsection{Scenario 2, Agents’ Diversity} \label{Agents’ Diversity}

The details of four different testing policies based on scenario 2 are discussed as below:

\subsubsection{Testing Policy 1: Red and Yellow}

Under this policy, the project manager gave agents with a belt rating of more than 1,500 points access to compete on the task. The first registration happened at time seven and had a failure prediction ratio around 1\%; the ratio rose to 11\% at time 14. At time 15 the first submission occurred and increased the failure prediction ratio to 18\%. The task submissions deadline continued until time 27, and the failure prediction increased to 83\%. On average, four agents were interested in this task and only one agent made a submission. Under this policy, the chance of receiving an unqualified submission was 80\%. Yellow belt agents made up 82\% of participants in this task while 18\% were red belt agents. Submission levels for yellow and red belt agents were 80\% and 20\% respectively.

\subsubsection{Testing Policy 2: Red, Yellow and Blue}

Under this policy, all agents with a rating above 1200 points were permitted to compete on the task. During time four, the first agent arrived, and the task failure prediction ratio was 5\%. It seems that the second arrived agent was more reliable than the first one as the failure prediction ratio reduced to 4\%. The first submissions occurred during time five and reduced the task failure prediction to 3\%. While registration still continued, after this point, only one more agent made a submission, which increased the failure prediction up to 40\%. On average, the task received ten registrations and three submissions under this policy, which provides a 60\% chance of task failure. Of the attracted agents, 62\% were blue belts, 31\% were yellow belts, and only 8\% were red belt agents. Interestingly, red belt agents did not make any submissions, while yellow belt agents made 33\% of submissions and the majority occurred by blue belt agents.

\subsubsection{Testing Policy 3: All but Gray}

In this policy, the project manager opened the task to all available agents with 900 rating points or above. The higher diversity of belt levels provided more competition starting from time three where the failure prediction level was 9\%. The first submission attempt occurred on the time four and decreased the failure prediction ratio to 5\%. While registration was continuing, the last submission for the task was received at time 12 and increased the failure prediction up to 40\%. Under this policy, the task attracted 12 registrants and a maximum of four submissions. There was a 73\% chance that the task received unqualified submissions.
Under this policy, 53\% of registrants were green belt agents, 26\% were blue belt agents, 16\% were yellow belt agents, and only 5\% were red belt agents. Against the expectation, more highly ranked agents did not make any submissions. While 50\% of submissions was made by green belt agents, blue belt agents submitted the other half.

\subsubsection{Testing Policy 4: All Welcome}

In this policy, all available agents in the platform were welcome to compete on the posted task. Simulation results revealed that the first agents arrived at time three and had a task failure prediction of 2\%. The second agent who arrived decreased the failure prediction to 1\%; however, the third agent arrival increased the failure prediction level. The first submission was made on the sixth run. After that submission, no further submissions were made during registration. This fact increased the task failure prediction to 93\%. On average, the task received 20 registrations and only one submission. The chance of receiving an unqualified submission under this policy was approximately 86\%.
This policy welcomed a wide diversity of available agents in the platform as competitors on this task. The competition level contained 44\% gray belt agents, 29\% green belt agents, 17\% blue belt agents, 7\% yellow belt agents, and 2\% red belt agents. Interestingly, while gray belt agents were the highest level of participants, they only made 18\% of the submissions. While green belt agents made 45\% of submissions, blue belt agents submitted 36\% of the time. Yellow and red belt agents did not submit for the tasks under this policy at all.

Table \ref{S-Senario2} summarizes agents’ participation in the different testing policies in Scenario 2.

\begin{table}[!ht]
\caption{Summary of Scenario 2} % title of Table
\centering % used for centering table
\begin{tabular}{p{0.4cm} p{1cm} p{1cm} p{0.7cm} p{0.7cm}p{0.7cm}p{0.7cm}}
\hline
\\
 \multicolumn{3}{c}{Scenario 2} & TP1 &  TP2 & TP3 & TP4\\ %[0.5ex] % inserts table
\hline%\hline % inserts single horizontal line
\\
\parbox[t]{2mm}{\multirow{1}{*}{\rotatebox[origin=c]{90}{Task Status}}}
&\multicolumn{2}{c}{Fail} & 18 & 22 & 25 & 23 \\
\\
&\multicolumn{2}{c}{Success} & 12 & 8 & 5 & 7 \\
\\
&\multicolumn{2}{c}{Failure Prediction} & 60\% & 73\% & 83\% & 77\% \\
\hline
\\
\parbox[t]{2mm}{\multirow{18}{*}{\rotatebox[origin=c]{90}{Agents' Participation}}}
&\multirow{3}{*}{Gray} 
& Reg &  &  &  & 44\% \\
\\
&& sub & &  &  & 18\%  \\
\\
&\multirow{3}{*}{Green} 
& Reg &  &  &  53\%& 29\% \\
\\
&& sub &  &  & 50\% & 45\%  \\
\\
&\multirow{3}{*}{Blue} 
& Reg &  & 62\% &  26\% & 17\% \\
\\
&& sub &  &  67\% & 50\% & 36\%  \\
\\
&\multirow{3}{*}{Yellow} 
& Reg &  82\% & 31\% &  16\% & 7\% \\
\\
&& sub &  80\% & 33\% & 0\% & 0\%  \\
\\
&\multirow{3}{*}{Red} 
& Reg &  18\% & 8\% &  5\% & 2\% \\
\\
&& sub &  20\% & 0\% & 0\% & 0\%  \\
\\
\hline
\label{S-Senario2}
\end{tabular}
\end{table}

\subsubsection{Impact of Scenario 2 on Task 8} \label{SCE2Task8}

\begin{figure}[ht!]
\centering
\includegraphics[width=0.9\columnwidth]{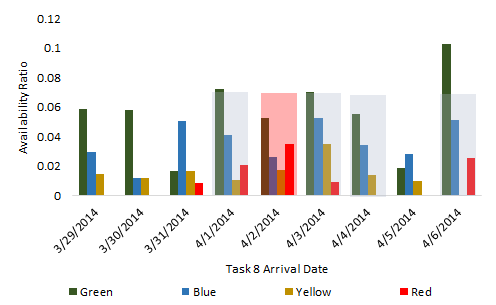}
\caption{Impact of Scenario 2 on Task 8}
\label{T8senario2}
\end{figure}

As displayed in Figure \ref{T8senario2}, the day that task 8 was posted in the platform, the availability of middle level agents (blue and green belts) was less than 5\%. The availability of mid-level agents was considerably higher on April 1\textsuperscript{st}, April 3\textsuperscript{rd}, April 4\textsuperscript{th}, and 6\textsuperscript{th}. Interestingly, according to empirical data on April 3\textsuperscript{rd} and April 4\textsuperscript{th}, the platform hosted the highest level of mid-level agents. It seems that if the task had been posted on either of those days, it might have had a higher chance of success. 
 According to the result of Scenario 2, if the task had been posted on April 6\textsuperscript{th}, the chances of failure would have decreased from 86\% to 73\% and the project schedule would have been 26 days shorter. Interestingly, based on available data, when the task was successfully completed as task 18, the average of mid-level agents’ availability was 7\%.

\end{document}